\documentclass{aa}  
\usepackage{graphicx}
\usepackage{natbib}
\bibpunct{(}{)}{;}{a}{}{,} 
\usepackage{txfonts}
\newcommand{\mysp}{}\def\mysp/{}
\newcommand{\allo}[3]{\ion{#1\mysp/}{#2}\ #3}
\newcommand{\forb}[3]{[\ion{#1\mysp/}{#2}]\ #3}
\newcommand{\sforb}[3]{\ion{#1\mysp/}{#2}]\ #3}

\newcommand{\allom}[3]{\ion{#1\mysp/}{#2}\ #3$\mu$m}

\newcommand{\alloa}[3]{\ion{#1\mysp/}{#2}\ #3\AA}
\newcommand{\dalloa}[4]{\ion{#1\mysp/}{#2}\ #3,#4\AA}
\newcommand{\forba}[3]{[\ion{#1\mysp/}{#2}]\ #3\AA}

\newcommand{\dforba}[4]{[\ion{#1\mysp/}{#2}]\ #3,#4\AA}
\newcommand{\rforba}[4]{[\ion{#1\mysp/}{#2}]\ #3/#4\AA}

\newcommand{\teff}{}\def\teff/{$\mathrm{T}_{\mathrm{eff}}$}
\newcommand{\hbeta}{}\def\hbeta/{H$\beta$}
\newcommand{\ibeta}{}\def\ibeta/{I$\beta$}
\newcommand{\halpha}{}\def\halpha/{H$\alpha$}
\newcommand{\cmage}{1400}
\newcommand{\cmlumi}{7685}
\newcommand{\cmlumilog}{3.88}
\newcommand{\cmcoeff}{1.20}
\newcommand{\cmbbtemp}{39.5}
\newcommand{\cmcmftemp}{36.7}
\newcommand{\cmebv}{0.26}
\newcommand{\cmrv}{3.6}

\begin{document}
   \title{A self-consistent stellar and 3D nebular model of planetary nebula IC418}

   \author{C.~Morisset  and L.~Georgiev}

   \offprints{C. Morisset}

   \institute{Instituto de Astronom\'{\i}a,
     Universidad Nacional Aut\'onoma de M\'exico\\
     Apdo. postal 70--264; Ciudad Universitaria;
     M\'exico D.F. 04510; M\'exico.\\
     \email{Chris.Morisset@gmail.com and Georgiev@astroscu.UNAM.mx}
}

   \date{Received May, 2009; accepted September, 2009}

 \abstract
{}
   {We present a coherent stellar and nebular model that reproduces observations of the planetary nebula IC418. We aim to test whether a stellar model found to provide an optimal description of the stellar observations is able to satisfactory ionize the nebula and reproduce the nebular observations, a finding that is by no mean evident. This allows us to determine all the physical parameters of both the star and the nebula, including chemical abundances and the distance.}
   {We used all the observational material available (FUSE, IUE, STIS and optical spectra) to constrain the stellar atmosphere model performed using the CMFGEN code. The photoionization model is developed by comparing solutions provides by Cloudy\_3D, with results from CTIO, Lick, SPM, IUE, and ISO spectra as well as HST images. The aperture sizes and positions of the different observations are taken into account. More than 140 model nebular emission lines are compared to the observed intensities. The distance is determined using evolutionary tracks.}
   {We reproduce all the observations for the star and the nebula. The 3D morphology of the gas distribution is determined. The effective temperature of the star is 36.7$\pm$0.5~kK. Its luminosity is 7700 L$_\odot$. No clumping factor is needed to reproduce the age-luminosity relation. We describe an original method for determining the distance of the nebula using evolutionary tracks. The distance of 1.25~kpc is found to be in very good agreement with recent determination using parallax method, and the age of the nebula is estimated to be 1400 years.
The chemical composition of both the star and the nebula are determined. Both are carbon-rich. The nebula exhibits evidence of the depletion of elements Mg, Si, S, Cl (0.5 dex lower than solar), and Fe (2.9 dex lower than solar), which is indicative of a depletion of these elements onto grains. }
   {We develop the first self-consistent stellar and nebular model of a planetary nebula that reproduces all the available observations ranging from IR to UV, showing that the combined approach to the modeling process leads to more restrictive constraints and, in principle, more trustworthy results.}


   \titlerunning{A full stellar and nebular model for IC418}
   \maketitle

             \section{Introduction}
\label{sec:intro}

Planetary nebulae (PNe) are the ultimate visible stage in the evolution of intermediate-mass stars. They consist of an extended shell photoionized by a central hot remnant (a white dwarf). The determination of the chemical composition of PNe is important for various reasons. It allows us to place important observational constraints on models nucleosynthesis of intermediate-mass stars \citep[e.g.,][]{2003A&A...409..619M}. It also allows one to trace the metallicity of the medium out of which the progenitor star was born, if using elements (S, Ar) whose abundances are not expected to be modified in the external layers of the planetary nebula (PN) progenitor during its evolution. Finally, it provides clues about the depletion of metals onto dust grains produced by the PN progenitor. 

Abundance determinations in planetary nebulae are obtained either by empirical methods, directly from the observed spectra, or through photoionization modeling. In principle, a photoionization model is expected to provide more reliable abundances, since these are obtained in a way that accounts for the temperature and ionization structure of the nebula. However, to provide reliable abundances, the model must reproduce the observed line ratios perfectly, which is by no means an easy task. The observing conditions must also be correctly simulated before comparing theoretical line ratios to observed ones. For example, if the observations are obtained through a slit that covers only a small fraction of the face of a nebula, one must compute the theoretical intensities as seen through such a slit.

The results from photoionization models are strongly dependent on the assumed spectral energy distribution (SED) of the ionizing star. The SED is often represented by a free set of parameters in the modeling procedure. However, in some cases, spectral observations of the central star exist, and allow one to perform a direct analysis of the properties of the stellar atmosphere (e.g., effective temperature, gravity, chemical composition) by comparing the observed stellar line shapes and intensities with those produced by a tailored stellar atmosphere model. 

While many detailed models of either planetary nebulae or PNe central stars have been constructed separately, there are only a few examples of studies combining both aspects \citep{2004MNRAS.354..558E,2007arXiv0709.2122W} and, for example, comparing the abundances determined for the star and the nebula.

In this paper, we present such a study for  \object{IC418}, an ellipsoidal PN of high surface brightness and low excitation. Its rather simple morphology combined with an impressive amount of observational material make this object an ideal test-case for a detailed combined model, associating up-to-date photoionization and atmosphere modeling codes.

This paper is structured as follows: a first section is devoted to the description of the observation material. The Section 2 describes the stellar and nebular modeling tools, as well as the stellar and  nebular modeling process and the interactions between the two. In the Sect.~3, we present the results of our best-fit model. In the Sect.~4, we discuss the expected accuracy of our results. In the Sect.~5, we discuss some implications of our findings, the most outstanding being related to the comparison between stellar and nebular abundances. 

             \section{Observations}
\label{sec:observations}
             \subsection{Stellar observations}
\label{sec:stellar-observations}

The spectrum of the central star were obtained from several sources. The data  from 900 to 1100 \AA\ were taken from FUSE observations obtained during the program P115 (PI J.M.Shull). A STIS/HST spectrum (dataset o52902) covers the range from 1190 to 1700 \AA\, and we finally combined the IUE spectra SWP37770, SWP37773, SWP37776, SWP37778, SWP37779, SWP37783, SWP37784, SWP37785, SWP37786, and SWP37792 to extend the UV spectrum to 1900 ~\AA. The existing high resolution LWP spectra are too noisy for detailed analysis. The photometric calibration of high resolution spectra were not very reliable, so we used two low resolution IUE spectra, SWP03178 and LWR03390, when comparing the observed and calculated stellar absolute flux. 

In addition to the UV spectrum, we used a high resolution (R $\sim$ 20000) optical spectrum of the central star obtained at the National Astronomical Observatory in San Pedro Martir (Mexico) using the REOSC echelle spectrograph. After the standard reduction performed with ESO-MIDAS \footnote{http://www.eso.org/sci/data-processing/software/esomidas/}, we extracted two nebular spectra in windows close to the star and of same size. These spectra are averaged, scaled, and subtracted from the stellar spectrum.
In that way, all but the strongest nebular lines were removed. Because of the wind velocity of $\sim$ 500 km/s, {H$\alpha$} is broader than the nebular lines and after subtraction of the nebula, its wings were clearly seen and we use them in the following analysis.

The central star of IC418 is a known irregular variable  with two characteristic timescales, one of a few hours and one of a few days \citep{1997A&A...320..125H}. The same authors report amplitudes of 0.1~mag in V  and a very small amplitude in B-V. \citet{1986RMxAA..13..119M} observed variations in  the equivalent width of \dalloa{C}{IV}{5808}{12} absorption lines on the shorter timescale.  The timescale of the rapid variability is shorter than the wind-crossing time and is filtered by it. \citet{1995A&AS..110..353P} analyzed several IUE spectra and concluded that the star does not show any detectable variability in the UV lines formed in the wind. These results means that the variability is principally related to the stellar photosphere. In the analysis presented below, we used mainly lines formed in the wind. Based on the small amplitude of the changes in both brightness and color and the insensitivity of the P Cyg profiles, we conclude that  the stellar variability does not affect significantly the model results presented in this paper.
             \subsection{Nebular observations}
\label{sec:nebular-observations}

Acquired in 1980-1981 with the Harvard spectral scanner, the CTIO 0.9m telescope observations were taken from \citet{1985PASP...97..397G}.
The Lick observatory observations (1991-1992) were reported by \citet{1994PASP..106..745H}. High spectral resolution spectra exhibiting emission line profiles were taken from \citet{1996A&A...309..907G}.

A deep optical echelle spectrum of IC 418 obtained in 2001-2002 at the 4m Blanco CTIO telescope is described by \citet{2004ApJ...615..323S}, who measured 807 emission lines of which 624 were solidly identified using EMILI software \citep{2003ApJS..149..157S}.

The HST WFPC observations were obtained from the database archive (proposal 8773, Hajian). Images in the \halpha/, \forba{N}{ii}{6584}, and \forba{O}{iii}{5007} filters were used.

IC 418 was observed in the UV by the International Ultraviolet (IUE) and the IR by the Infrared Space Observatory (ISO). We take the  line intensities corrected for extinction from \citet{2004A&A...423..593P}.

The UV to optical normalization was obtained following \citet{2004A&A...423..593P}, using the \rforba{O}{ii}{2471}{7325} line ratio. We use the theoretical value given by the model (namely 0.75), taking into account the difference between the aperture sizes of the optical and UV observations. The optical values used in this calibration are those observed by \citet{2004ApJ...615..323S}. At the end of the convergence process, we check that the absolute fluxes are reproduced.

To connect the intensities observed by ISO to the optical intensities, we use the Balmer and Brackett hydrogen lines. These lines were observed with the ISO SWS spectrometer, while the lines of wavelength longer than 50 $\mu$m were observed by the LWS spectrograph. Since both spectrometers have aperture sizes that are larger than the entire nebula, no differential aperture corrections are needed. This is confirmed by the continuum emission (supposed to arise from the external dust around the ionized nebula) being of very similar intensity when observed by both spectrometers in a common wavelength range (around 45 $\mu$m).

Notice that the absolute fluxes predicted by the model for \forba{O}{ii}{2471}, the Brackett, and the Balmer lines are in a very good agreement with the observations, meaning that the calibrations we use to connect the IR, optical, and UV observations are close to the absolute calibration. The sizes and positions of the apertures used by \citet{2004ApJ...615..323S} and \citet{1994PASP..106..745H} are taken into account, as described in Sect.~\ref{sec:nebular-model}.

\citet{1985PASP...97..397G} report observations in the lines of \forba{Ne}{v}{3425}, \forba{Ar}{v}{7006} and \alloa{He}{ii}{4686}. We use the atmosphere model to verify whether the \alloa{He}{ii}{4686} line could be of stellar origin. The \forba{Ne}{v}{3425} and \forba{Ar}{v}{7006} are not emitted by either the star or the nebula.

             \section{The models}
             \subsection{Stellar model}
\label{sec:CMFGEN}

One of the major problems in compiling any photoionization model of a PN, apart from the density distribution (discussed in Sect.~\ref{sec:density-distribution}), is the description of the ionizing spectrum. By definition, the photons capable of ionizing the gas are never observed, and the simplest hypothesis involves use of a black-body distribution, defined only by an effective temperature and scaled by a luminosity. In an effort to understand the ionization state of the nebula more clearly and to resolve some discrepancies between the model predictions and the observations, we use a more realistic description of the ionizing flux by modelling in detail the atmosphere of the ionizing star.

The stellar atmosphere model is obtained using the CMFGEN code from \citet{HM98}. CMFGEN is a non-LTE line blanketed code designed for spectral analysis of stars with winds. The radiative transfer equation is solved in the comoving frame by assuming statistical and radiative equilibrium. The code is capable of treating a large number of ions with an enormous number of individual levels. As a result, the emitted spectrum is reliable across a wide range of frequencies including the flux above 13.6 eV that is responsible for the ionization of the nebula.

In contrast to other models of the star, the effective temperature is fixed not by stellar lines but using the ionization balance of the nebula. Once the ionization of the nebula is reproduced, we then check the temperature structure of the wind using the lines of \ion{N}{III}, \ion{N}{IV}, and \ion{N}{V}, and \ion{C}{III} and \ion{C}{IV}.  In that way, the stellar model is consistent with the nebular model. It reproduces correctly the stellar flux bluer than the H ionization edge, which is not seen by the observer but strongly affects the nebula. 

The density structure of the wind is set by an ad-hoc velocity law in the form \citep{2003ApJ...588.1039H}
\begin{equation}
v(r) = \frac{V_{phot} + \left(V_\infty - V_{phot}\right) \left(1-R_{star}/r \right)^\beta} {1 + \left(V_{phot}/V_{core}-1\right)\exp{\left(-\left(r-R_{star}\right)/h\right)}},
\end{equation}
where $V_{core}$ represents the depth to which the model extends and $V_{phot}$ governs the match between the hydrostatic atmosphere and the wind. The scale factor $h$ is defined as 
\begin{equation}
h = \frac{k T \left(1-\gamma\right)}{g \left(1-\Gamma\right) \mu m_\mu },
\end{equation}
where $\gamma$ is the mean number of electrons per atom, $\Gamma = g_{rad}/g$ is the ratio of the radiation pressure to the gravity acceleration, $\mu$ is the mean atomic weight, and $m_\mu$ is the atomic mass unit.  

The value V$_\infty$=400 km/s is set by the blue wings of {\it all} UV P Cyg profiles. For the \dalloa{C}{IV}{1548}{50} lines, we fit the bluest part of the saturated absorption component (sometime called $V_{black}$) instead of the point at which the blue wing reaches the continuum called $V_{edge}$, which is usually affected by the velocity dispersion in the wind and can be higher than $V_{black}$ by 200-400 km/s. The model with the derived value of V$_\infty$, although lower (by a factor of two) than the previous studies \citep{1988A&A...190..113M,1997IAUS..180...64K,2004A&A...419.1111P}, reproduces in detail a large number of lines. We attribute the discrepancy with other models to the higher quality of the spectra (mainly FUSE and HST/STIS) used in this work, which were unavailable at the time of previous studies. 

\begin{figure}
  \centering
    \includegraphics[width=8.0cm]{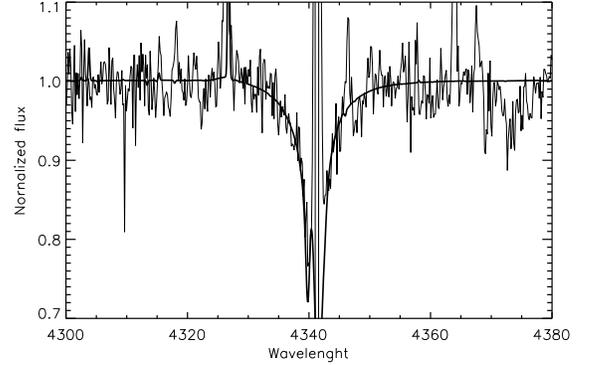}
   \caption{Comparison between observed wings of H$\gamma$ (thin line) and the model with 
   log g = 3.55 (thick line)}
  \label{fig:h_gamma}
\end{figure}

\begin{figure*}
  \centering
    \includegraphics[width=16.cm]{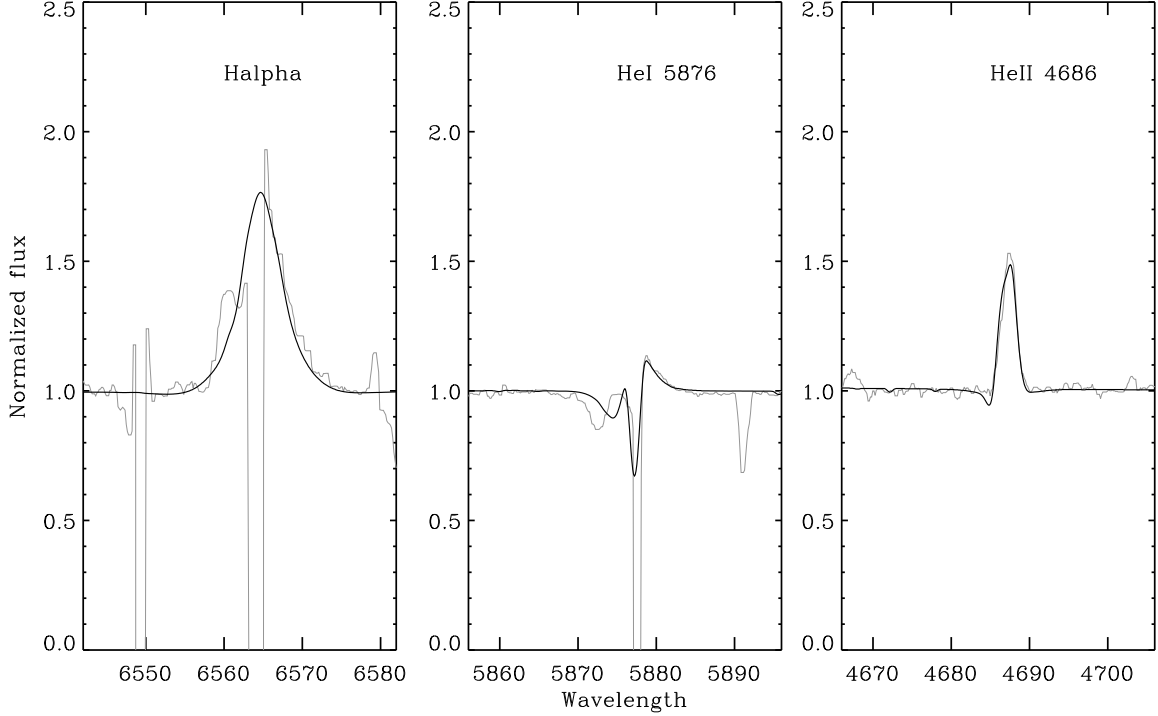}
   \caption{The fit to the \halpha/, \alloa{He}{I}{5876},  and \alloa{He}{II}{4686} lines determines the mass-loss rate and the helium abundance. Observations are plotted in gray, the model is in black. The absorption component blueward of \alloa{He}{I}{5876} is absent in a spectrum taken 3 days later. The lines are saturated in the spectrum and their subtraction is incorrect. The nebular subtraction removes the weak lines correctly so we expect that the oversubtraction of the saturated \halpha/ and \ion{N}{ii} line will not affect the  wings of the stellar \halpha/, which were used in the analysis.}
  \label{fig:he_fig}
\end{figure*}

When the terminal velocity has been fixed, we determine the value of $\beta=3.0$ by fitting the width of \alloa{He}{II}{4686} (Fig.~\ref{fig:he_fig}). The value of $h$ (and therefore log g) is determined by fitting the wings of H$_\gamma$. Figure~\ref{fig:h_gamma} shows the comparison between our model with $h$ = 0.025R$_\odot$, which corresponds to log g = 3.55, and the observed H$_\gamma$ wings. We use a turbulent velocity of 10 km/s during the modeling and then broaden the calculated spectrum with a 30 km/s rotational velocity to fit the absorption lines observed in the UV spectrum. Table~\ref{tab:param} summarize the parameters obtained.

After fixing the wind parameters, we determine the abundances of the main elements. The mass loss rate and the He/H ratio is determined simultaneously from the fit to the \halpha/, \alloa{He}{I}{5876},  and  \alloa{He}{II}{4686} lines (Fig.~\ref{fig:he_fig}). It was shown \citep{2007MNRAS.382..299P} that \dalloa{P}{V}{1117}{28} is sensitive to the wind clumping factor. The assumed clumping factor, f = 0.1, reproduces this line profile so well that we do not adjust it during the fitting procedure.
The carbon abundance is measured from the fit of \alloa{C}{III}{1178}, \alloa{C}{III}{1247}, and \alloa{C}{IV}{1169} lines (Fig~\ref{fig:carb_fig}, online material). 
The nitrogen abundance is determined from the \alloa{N}{III}{1750} and \alloa{N}{IV}{1718} lines (Fig.~\ref{fig:nit_fig}, online material). 
The Oxygen abundance is set by \dalloa{O}{IV}{1339}{41} and \alloa{O}{III}{1150}, \alloa{O}{III}{1151}, \alloa{O}{III}{1155} lines (Fig.~\ref{fig:oxy_fig}, online material). The ratios of the lines for different ionization stages of He, C, and N agree with the temperature determined by the nebular response. However, the observed lines of the three ionization stages of oxygen imply that the temperature is $\sim$ 2000K higher. On the other hand, the FUSE spectrum of IC418 shows strong \dalloa{O}{VI}{1031}{37} emission (Fig.~\ref{fig:oxy_fig}), which is indicative of a super-ionization effect probably affecting all the oxygen ionization stages \citep{1994ApJ...437..351M}. 
The oxygen abundance is assumed to be a compromise between that related to the \ion{O}{iii} and \ion{O}{iv} lines, because the superionization effect is understood to have less influence on these ions. In addition, the oxygen abundance affects the intensity of \alloa{He}{II}{4686} indirectly. The assumed oxygen abundance is consistent with \alloa{He}{II}{4686} intensity, and we therefore assign a higher uncertainty of the oxygen abundance. 

The effect of superionization on the wind structure is unclear. \citet{1994ApJ...437..351M} showed that the presence of X-ray emission can alter all the ionization stages of most of the important ions. On the other hand, existing X-ray observations of IC418 do not detect a point source. We know that some superionization exists only by the presence of \dalloa{O}{VI}{1031}{37}. This problem has no simple solution, and we leave its detailed study to a subsequent paper. The presence of additional ionization, which is not taken into account in the model, infers that all the abundances obtained from the model into lower limits.  

The phosphorus abundance is determined from \dalloa{P}{V}{1117}{21}, the silicon abundance from the \dalloa{Si}{IV}{1392}{1402} and \alloa{Si}{IV}{1122} lines, the sulfur abundance from \alloa{S}{V}{1502} (Fig.~\ref{fig:phos_fig}) and several other FUV lines and, finally, the argon abundance is determined from the \alloa{Ar}{V}{1346} line (Fig.~\ref{fig:oxy_fig}). The iron abundance is measured by fitting the iron lines between 1250 and 1500 \AA. The strong lines are usually fitted reasonably well, but there are a large number of relatively weak lines, some of which are overestimated and some underestimated. We attribute these discrepancies to the incorrect oscillator strengths. The neon abundance cannot be determined because of the absence of observable lines. We define its abundance to be its solar value. The abundance of each element is first determined using the lines mentioned above and then checked by comparison with all other lines present in the observed spectra.
The parameters of the model that reproduces both the nebular and the stellar spectra are presented in Sect.~\ref{sec:results}.

We note that one cannot obtain a unique solution for the stellar luminosity by fitting only the normalized stellar spectrum. The parameters describing the star are not independent. If the star is described as an opaque nucleus with radius $R_*$ and temperature $T_*$ embedded in the wind then the luminosity is $L = 4 \pi \sigma R_*^2 T_*^4$. The effective temperature $T_{eff}$ is defined to be the wind temperature at a radius where the Rosseland optical depth, $\tau_{Ross}$, reaches $2/3$. 
It is known \citep{1989A&A...210..236S}, that models of the same temperature $T_*$ and same transformed radius
\begin{equation}
R_t = R_* \left(\frac{V_\infty/2500}{\dot M/10^{-8}}\right)^\frac{2}{3}, \nonumber
\end{equation}
have very similar emission-line spectra. Substituting for the radius $R_*$ one obtains a scaling rule for the mass-loss rate of
\begin{equation}
\frac{\dot M_1}{\dot M_2} = \left(\frac{L_1}{L_2}\right)^{\frac{3}{4}}, \nonumber
\end{equation}
which implies means that an increase in the luminosity $L$ can be compensated by an increase in the radius $R_*$ and an increase in $\dot M$. 
There is no easy way of removing this degeneracy except by knowing the distance. In Sect.~\ref{sec:distance}, we discuss one way of deriving the distance.
We note that we also take into account the X-ray emission in the modeling of the ionizing SED. This component does not have an important effect on the determination of the main parameters of either the star or the nebula, thus this part is discussed in Appendix~\ref{sec:xray}.

             \subsection{Nebular model}
\label{sec:density-distribution}
\label{sec:nebular-model}

The models presented in this paper are obtained using Cloudy\_3D (hereafter C3D), described in \citet{2006IAUS..234..467M}. We choose to use a 3D description of the object to take into account the different sizes and positions of the apertures used by various authors, whose data we compare with the model.
C3D is an IDL (ITT) library of routines based on Cloudy \citep{1998PASP..110..761F}, which produces a pseudo-3D photoionization model by combining several radial 1D models with the code Cloudy. In the present paper, we use version c07.02.01 of Cloudy. Each run of Cloudy corresponds to a given radial direction from the star, the input parameters being changed to reflect the geometry (e.g., inner radius, radial position of the two shells, hydrogen density). The 3D nebula is then obtained by interpolation of the results of the 1D runs, leading to a cube of physical parameters (electron temperature, ionic abundances) and emissivities of the lines of interest. More details about C3D can be found in \citet{2006IAUS..234..467M}.

We note that the apparent granularity of the nebula as seen on HST images consists only of variations of 10\% in the surface brightness, which could be due to variations of 5\% in the density.  These small variations have no effect on the determination of the global properties of the nebula and we neglect them by assuming a smooth density distribution.

From the images of IC418, we assume that the nebula is axisymmetric, with a 3D distribution of ionized gas that follows an ellipsoidal shape. The nebula should not have an inclination angle (that is, its axis of symmetry is in the plane of the sky). It is also supposed that the nebula is radiation-bounded in all the directions. 
This rather simple morphology allows us to obtain complete models with only 4 runs of Cloudy covering polar angles between 0 and 90~deg. Because of the polar-axial and equatorial-planar symmetries of the nebula, only 1/8 of the entire object need be computed by interpolation. We use a cube of 40$^3$ pixels, the entire nebula being simulated by merging the same cube after adequate rotations, leading to a 80$^3$ pixel cube.

\begin{figure}
  \centering
    \includegraphics[width=7.cm]{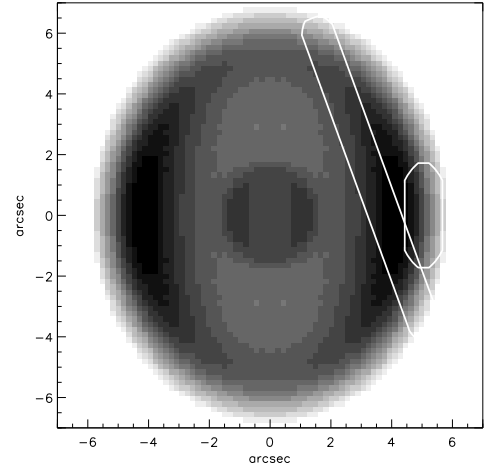}
  
  \caption{Apertures used to reproduce observations by \citet{2004ApJ...615..323S} (left one) and \citet{1994PASP..106..745H} (right one), superimposed on the \hbeta/ model image.}
  \label{fig:apertu}
\end{figure}

Emission-line maps are then obtained by summing the emissivity cube in a given direction (small axis in our present case). 
To compare with the observed line intensities, the slit sizes and positions are also simulated.
Some of the optical observations were obtained with apertures smaller than the entire projected size of the object \citep{1994PASP..106..745H,2004ApJ...615..323S}. In the following, we apply the same aperture sizes, orientations, and positions to the emission-line intensity maps when comparing with these author's line intensities, while the full object is taken into account for the UV and IR intensities. Figure~\ref{fig:apertu} shows the apertures used. We note that from the models described below, we found that only 10\% (3.5\%) of the \hbeta/ flux is intercepted by the optical aperture used by \citet{2004ApJ...615..323S} \citep[][resp.]{1994PASP..106..745H}.

The free parameters needed to fully describe a model and compare it to observations are: the 3D density distribution, the angular size of the nebula (or its distance), the nebular abundances, the ionizing photon distribution, and the number of ionizing photons emitted per unit time (Q$_0$). 
The processes that determine the adopted values of these parameters are described in the following subsections.

The morphology of IC418 was obtained by trial and error, using the HST images of \hbeta/, \forba{O}{iii}{5007}, and \forba{N}{ii}{6584} and density diagnostics (11 line ratios).

The main goal in the determination of a somewhat complex density distribution is to reproduce the central enhanced emission from \forba{O}{iii}{5007}. This is reached by using a double ellipsoidal shell morphology, where the density is adjusted to reproduce the emission-line diagnostics. The density distribution cannot be recovered precisely without a high resolution density map, as for example given by a \rforba{S}{ii}{4068}{4076} map. Nevertheless, the density distribution obtained here is constrained sufficiently by the overall density diagnostics and the surface brightness maps. 

The adopted density distribution consists of two ellipsoidal shells of distinct densities, sizes, and widths. Its analytical description in a spherical coordinate system is:

\begin{equation}
  \begin{array}{rl}
    n_H(R<R_{in}) = & 0, \\
    n_H(R>R_{in}, \theta) = & n_0 + \\
    & n_1(\theta) \times \exp(-(\frac{R-R_1(\theta)}{\sigma_1})^2) + \\
    & n_2(\theta) \times \exp(-(\frac{R-R_2(\theta)}{\sigma_2})^2), 
  \end{array}
  \label{eq:morf}
\end{equation}

\noindent where the angular dependencies of the parameters are as follows: 
\noindent $R_1(\theta) = R_1 \times F_{ell-1}(\theta), R_2(\theta) = R_2 \times F_{ell-2}(\theta), n_1(\theta) = n_1 / F_{ell-1}(\theta)$ and $n_2(\theta) = n_2 / F_{ell-2}(\theta)$, where $\theta$ is the polar angle (set to be 0 in the equatorial direction), and $F_{ell}(\theta)$ is an ellipsoid magnification factor defined by:

$$ F_{ell-i}(\theta) = a_i / \sqrt{(\sin(\theta))^2 + (a_i \times \cos(\theta))^2}, $$

\noindent where $a_i$ is an adimensional coefficient, set to be 1.3 (i=1) and 1.5 (i=2) for the inner and outer ellipse respectively. The numerical values of these parameters for the best-fit solution that we obtain are given in Table~\ref{tab:param}.

\begin{figure}
  \centering
    \includegraphics[width=9.cm]{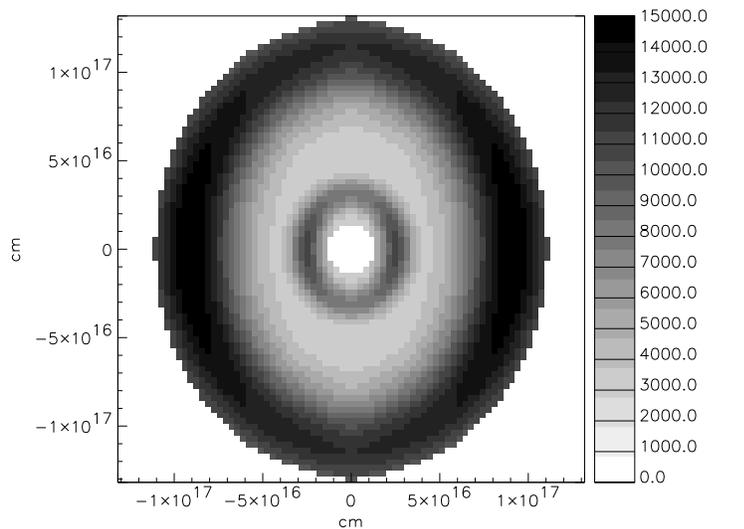}
  \caption{Cut in the density distribution through a plane including the polar axis, units in cm, gray-scale in hydrogen atom/cm$^3$.}
  \label{fig:dens}
\end{figure}

Figure~\ref{fig:dens} shows the hydrogen density distribution in a plane containing the polar axis of symmetry. The full density distribution is obtained by rotation about the polar axis. We will use this density distribution for the model presented in the following sections.

\begin{figure*}
  \centering
    \includegraphics[width=16.cm]{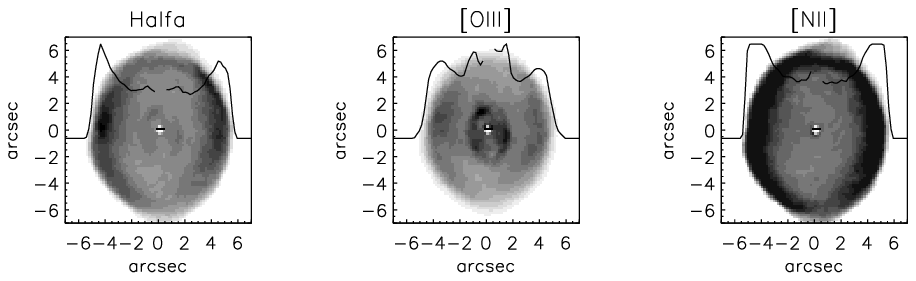}
    \includegraphics[width=16.cm]{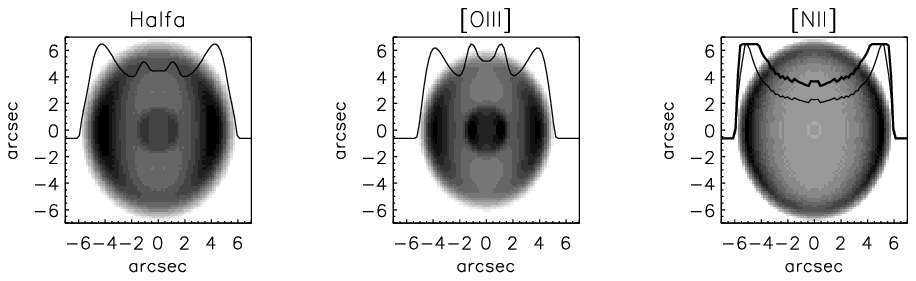}
   \caption{We compare the HST images (top) and the model images (bottom) for 3 emission lines. Units are in arcsec. Gray values are on a linear arbitrary scale. The lines are Y=0 arcsec cuts in the surface brightness maps.
  }
  \label{fig:hst}\label{fig:model}
\end{figure*}

Figure~\ref{fig:hst} shows \hbeta/, \forba{O}{iii}{5007}, and \forba{N}{ii}{6584} images of the nebula, and the ratios of these lines, for both the HST observations and the model, respectively. The curves are Y=0 arcsec cuts in the surface brightness, passing through the central point. In the case of the HST images, the central star emission has been removed to allow a clearer comparison between the model and the observed images. The global shape of the nebula is reproduced well by our model. 

The flat top of the \forba{N}{ii}{6584} image is due to saturation effect in the observation. We artificially saturate the model image by taking the same ratio of the mean value at the center of the image (without the star) and the saturation value. The resulting saturated profile is shown with a thicker line, reproducing the observed profile.

When the density distribution is determined, we run models with the stellar SED determined in Sec.~\ref{sec:CMFGEN}. The abundances of the elements are then determined by fitting the most important lines of all the observable ions of each element. A total of 140 emission lines are used for the model, most of them being pure predictions (not used for the convergence process). The complete process of convergence is achieved by fine tuning the effective temperature until we successfully reproduce the ionization state of the nebula, once the \hbeta/ absolute flux is well reproduced.

The ISO observations show a continuum emission, which is commonly associated with dust. The emission reaches a maximum value of 400~Jy close to 35~$\mu$m. It is beyond the scope of this paper to reproduce the exact shape of the dust emission. Nevertheless, we do include dust in the photoionization model to take into account its absorption when fitting the stellar continuum and determining the interstellar extinction. We check that the dust emission is globally of the order of what is observed.

             \section{Presentation of our best-fit model}
\label{sec:results}

             \subsection{Parameters of the nebular and stellar models}
\label{sec:parameters}

The parameters of the adopted model are described in Table~\ref{tab:param}. Table~\ref{tab:abund} lists the abundances of the adopted models (stellar and nebular) and values from the literature (for the nebula).  These abundances by mass (for the model and the literature data) are also presented in Fig.~\ref{fig:abund} (online only). We note that our values for C/H, N/H, and O/H are comparable to the highest values obtained previously, which also correspond to the more recent determinations \citep[namely ][]{2004A&A...423..593P,2007ApJS..169...37S}.

\begin{table}
  \caption{Parameters of the model.}
  \label{tab:param}
  \centering
  \begin{tabular}{|l|r|r|}
    \hline
    Parameter & Unit & Adopted model \\
    \hline
    \noalign{\smallskip}    
    Teff & kK  & \cmcmftemp \\
    log(Q$_0$) & s$^{-1}$      &47.5\\
    log(L$ $) & L$_{\odot}$ & \cmlumilog \\
    log(g) &  & 3.55 \\
    V$_\infty$ & km.s$^{-1}$ & 450 \\
    Mdot & 10$^{-8}$M$_\odot.yr^{-1}$ & 3.6 \\
    Nucleus Mass & M$_\odot$ & 0.6 \\
    Teff(X) & kK & 290 \\
    log(L(X)) & L$_{\odot}$ &  0.5 \\
\noalign{\smallskip}    \hline \noalign{\smallskip}
    
    $n_0$ &cm$^{-3}$ & 2700\\
    $n_1$ & cm$^{-3}$ & 7500\\
    $n_2$& cm$^{-3}$ & 11700\\
    $<n>$& cm$^{-3}$ & 9300\\    

    log($R_{in} $)& cm &16.09\\
    log($R_1 $)& cm &16.16\\
    log($R_2 $)& cm &16.96\\
    
    log($\sigma_1 $)& cm &16.\\
    log($\sigma_2 $)& cm &16.5\\
    
    $ff_C$  &   & 1.   \\  
\noalign{\smallskip}    \hline\noalign{\smallskip}
    Distance$ $ & kpc & 1.26 \\
    Age$ $ & a & \cmage \\
    H$^+$ mass & M$_{\odot}$ & 0.06 \\
\noalign{\smallskip}    \hline\noalign{\smallskip}
    T0   & K  & 8575 \\
    t$^2$  &    & 0.002 \\
    T(O$^+$) & K & 8690\\
    T(O$^{++}$) & K & 8375\\
    T(N$^+$) & K & 8810\\
    T(N$^{++}$) & K & 8375\\
\noalign{\smallskip}    \hline
    
  \end{tabular}
\end{table}

\begin{table*}\centering
  \caption{Abundances from the literature and for the adopted model} 
  \label{tab:abund}
  \begin{tabular}{|l|r|r|r|r|r|r|r|r|r|r|r|r|}
    \hline \noalign{\smallskip}
                           & He    & C     & N     & O     &  Ne   &   Mg  & Si    &   P   & S     & Cl    & Ar    & Fe \\
    \noalign{\smallskip} \hline \noalign{\smallskip}
    \multicolumn{13}{|c|}{BY NUMBER in log} \\
    \noalign{\smallskip} \hline \noalign{\smallskip}
 Best Stellar Model&   -0.60 &    -3.30 &    -4.24 &    -3.42 &    &    &    -4.29 &    -7.11 &    -5.36 &    &    -5.29 &    -4.74 \\
 Best Nebular Model&   -0.92 &    -3.10 &    -4.00 &    -3.40 &    -4.00 &    -4.95 &    -4.90 &    &    -5.35 &    -7.00 &    -5.80 &    -7.40 \\
    \noalign{\smallskip} \hline \noalign{\smallskip}
    \multicolumn{13}{|c|}{LITERATURE NEBULAR ABUNDANCES (BY NUMBER in log)} \\
    \noalign{\smallskip} \hline \noalign{\smallskip}
    \citet{2007ApJS..169...37S} Model A        &-0.975 &-3.22&-3.97&-3.29&-3.99&      &      &     &-5.40&-7.07&-5.79&     \\
    \citet{2007ApJS..169...37S} Model B        &-0.850 &-3.32&-3.97&-3.43&-3.75&      &      &     &-5.35&-7.11&-5.81&     \\
    \protect{\citet{2004A&A...423..593P}}      &-1.142 &-3.21&-4.02&-3.46&-4.05&      &      &     &-5.36&-6.92&-5.74&     \\
    \citet{2000ApJ...531..928H}&-1.066 &-3.66&-4.07&-3.81&-5.02&      &     &     &          & &     &     \\
    \citet{1994PASP..106..745H} Model&-1.155 &-3.52&-4.15&-3.74&-5.52&-5.55 &-6.00&      &-5.60&-7.00&-6.26&     \\
    \citet{1994PASP..106..745H} ICF  & -1.155&-3.57&-4.08&-3.67&-4.22&-5.16 &     &      &-5.56&-7.11&-5.72& \\
    \protect{\citet{1980Ap&SS..67..349A}}&-0.920 &-3.22&-4.22&-3.37&-4.40&      &      &     &-5.38&-6.85&-5.34&     \\
    \citet{1983ApJS...51..211A}&-1.030 &-3.21&-4.13&-3.36&-4.13&      &     &      &-5.38&-7.06&-5.64&     \\
    \citet{1981ApJ...249..592B}&       &     &     &     &-4.12&      &     &      &-5.21&     &-5.74&     \\
    \citet{1973ApJ...180..817B}&-0.796 &-3.70&-4.40&-3.60&-4.50&      &     &      &-5.00&-6.82&-5.30&     \\
    \noalign{\smallskip} \hline \noalign{\smallskip}
    Solar \citep{2005ASPC..336...25A}&-1.070 &-3.61&-4.22&-3.34&-4.16&-4.47 &-4.49& -7.00&-4.86&-6.50&-5.82&-4.55\\
    \noalign{\smallskip} \hline \noalign{\smallskip}
    \multicolumn{13}{|c|}{BY MASS in log} \\
    \noalign{\smallskip} \hline \noalign{\smallskip}
 Best-fit stellar model&   -0.30 &    -2.52 &    -3.40 &    -2.52 &    &   &    -3.15 &    -5.92 &    -4.16 &    &    -3.99 &    -3.30 \\ 
 Best-fit nebular model&   -0.49 &    -2.20 &    -3.03 &    -2.37 &    -2.88 &    -3.75 &    -3.63 &    &    -4.02 &    -5.63 &    -4.37 &    -5.83 \\ 
    Solar \citep{2005ASPC..336...25A}&     -0.60  & -2.66 &  -3.21 & -2.27 & -2.99 & -3.22 & -3.17 & -5.28 & -3.49 & -5.09 & -4.35 & -2.93 \\ 
    \noalign{\smallskip} \hline \noalign{\smallskip}
  \end{tabular}
\end{table*}

\subsection{Comparing the stellar parameters with previous studies}
\label{sec:comp-stell-param}

Stellar models of the central star of IC418 are available in the literature from \citet{1988A&A...190..113M,1997IAUS..180...64K,2004A&A...419.1111P}. The stellar parameters obtained by these authors and in this work are summarize in Table~\ref{tab:stel_param}. In general, the parameters obtained in this paper are similar to those obtained by other authors although the values presented here have the advantage of being consistent with the nebula and evolutionary status of the star. Our value of V$_\infty$ reproduces most of the features in the observed spectrum, especially the FUV lines that were not available at the time of previous modeling.  In general, one can conclude that the parameters obtained here are compatible with the previous values and the main contribution of this paper is the more reliable and consistent values and the chemical composition of the star.

\begin{table*}\centering
  \caption{Stellar parameters from the literature and for the adopted model} 
  \label{tab:stel_param}
  \begin{tabular}{|l|r|r|r|r|r|c|r|r|}
    \hline \noalign{\smallskip}
   Reference               & log(L/L$_\odot$)   & T(kK) & log g & M/M$_\odot$ & d(kpc) & $\dot  M$ (10$^{-7}$M$_\odot$/yr)& V$_\infty$ & R/R$_\odot$ \\
    \noalign{\smallskip} \hline \noalign{\smallskip}
    \protect{\citet{1988A&A...190..113M}}& 4.2  &  36.  & 3.3 & 0.77 & 2.0  &      &      & 3.25 \\
    \protect{\citet{1997IAUS..180...64K}}& 4.3  &  37   & 3.3 & 0.89 &      & 2.6  &  700 & 3.5  \\
    \protect{\citet{2004A&A...419.1111P}}& 4.2  &  39   & 3.7 & 1.33 & 2.0  & 0.72 &  800 & 2.7  \\
    This work                            & 3.88 &  36.7 & 3.55 & 0.60 & 1.26 & 1.1$^a$ &  450 & 2.17 \\
    \noalign{\smallskip} \hline \noalign{\smallskip}
  \end{tabular}
  
    a) Value of smooth wind which correspond to 3.6$\times$10$^{-8}$  M$_\odot$/yr with clumping factor f=0.1.
\end{table*}

             \subsection{Defining tolerance and quality factor $\kappa({\rm O})$ to compare the nebular model and the observations}
\label{sec:definning-tolerance}

To compare the results of the model with observations, we define errors in the observed intensities, $\frac{\Delta I}{I}$, as the following: 10\% for lines brighter than 0.1 \hbeta/, 20\% for lines of strength between 0.01 and 0.1 \hbeta/, and 30\% for lines lower than 0.01 \hbeta/. To take into account the lowest signal-to-noise ratio of the IUE and ISO observations and the uncertainties in the connection between the optical and the IR and UV domains, we add 20\% to the previously determined errors of the IR and UV lines. We also add 10\% to the optical lines obtained by \citet{1985PASP...97..397G,1994PASP..106..745H}. We note that these errors also aim to take into account the numerical and modeling uncertainties due to the limited complexity of the morphology that we adopt, the errors in the atomic data that we use, and the effects of the position and size of the apertures that we use (see the comments about rotating apertures during the observations by \citet{1994PASP..106..745H}).

A quality factor of the fit to an observable is defined by 
\begin {equation} 
  \kappa({\rm O}) = \frac{{\rm log}( {\rm O}_{\rm mod}) - {\rm log}( {\rm O}_{\rm obs})}{\tau({\rm O})},
\end {equation} 
where ${\rm O}_{\rm mod}$ is the value returned by the model, ${\rm O}_{\rm obs}$ is the
observed value, and $\tau({\rm O})$ is the accepted tolerance in dex of this observable. 
We define the tolerance to be $\tau({\rm O}) = {\rm log} (1+\frac{\Delta I}{I})$. 
When the absolute value of $\kappa({\rm O})$ is lower than~1, the model value is within the adopted tolerance of the observed value. The value of $\kappa({\rm O})$ is also used to analyze the diagnostic line ratios.

             \subsection{Line ratios and absolute fluxes}
\label{sec:comparing-model-with}

\begin{table*}
  \caption{Results of the Teff=\cmcmftemp~kK CMFGEN model, see the text for details.} \label{tab:res1}
\centering
{\tiny   \begin{tabular}{|l|r|r|r|r||l|r|r|r|r||l|r|r|r|r|}
 \hline
    \multicolumn{1}{|c|}{\textbf{I}} 
    & \multicolumn{1}{c|}{\textbf{II}}
    & \multicolumn{1}{c|}{\textbf{III}}
    & \multicolumn{1}{c|}{\textbf{IV}}
    & \multicolumn{1}{c||}{\textbf{V}}
    & \multicolumn{1}{|c|}{\textbf{I}} 
    & \multicolumn{1}{c|}{\textbf{II}}
    & \multicolumn{1}{c|}{\textbf{III}}
    & \multicolumn{1}{c|}{\textbf{IV}}
    & \multicolumn{1}{c||}{\textbf{V}}
    & \multicolumn{1}{c|}{\textbf{I}} 
    & \multicolumn{1}{c|}{\textbf{II}}
    & \multicolumn{1}{c|}{\textbf{III}}
    & \multicolumn{1}{c|}{\textbf{IV}}
    & \multicolumn{1}{c|}{\textbf{V}}\\
    \hline
            \allo{H}{I  }{3835}&0& {\bf      9.490}&     7.841&{\it      -1.05} &        \allo{N}{I  }{8212}$^a$ &0& {\bf      0.142}&     0.002&{\it     -16.64} &             \forb{S}{II }{4070}&0& {\bf      1.780}&     2.313&{\it       1.44}\\
             \allo{H}{I  }{3835}&3& {\bf      6.950}&     7.829&{\it       0.45} &            \sforb{N}{III}{1750}&2& {\bf      0.584}&     0.637&{\it       0.21} &             \forb{S}{II }{4070}&3& {\bf      2.440}&     3.209&{\it       1.04}\\
             \allo{H}{I  }{3970}&0& {\bf     16.900}&    16.511&{\it      -0.24} &             \forb{N}{III}{57.2}&1& {\bf      3.323}&     2.722&{\it      -0.59} &             \forb{S}{II }{4078}&0& {\bf      0.765}&     0.750&{\it      -0.07}\\
             \allo{H}{I  }{3970}&3& {\bf     15.980}&    16.501&{\it       0.18} &             \allo{N}{II }{4041}&0& {\bf      0.012}&     0.018&{\it       1.43} &             \forb{S}{II }{4078}&3& {\bf      0.880}&     1.041&{\it       0.50}\\
             \allo{H}{I  }{4102}&0& {\bf     24.800}&    26.575&{\it       0.73} &             \allo{N}{II }{4176}&0& {\bf      0.006}&     0.003&{\it      -3.43} &             \forb{S}{II }{6716}&0& {\bf      2.080}&     1.863&{\it      -0.60}\\
             \allo{H}{I  }{4102}&3& {\bf     25.240}&    26.580&{\it       0.28} &             \allo{N}{II }{4239}&0& {\bf      0.013}&     0.012&{\it      -0.17} &             \forb{S}{II }{6716}&3& {\bf      3.590}&     2.501&{\it      -1.38}\\
             \allo{H}{I  }{4340}&0& {\bf     44.800}&    47.326&{\it       0.58} &             \allo{N}{II }{4435}&0& {\bf      0.003}&     0.008&{\it       3.81} &             \forb{S}{II }{6716}&4& {\bf      2.090}&     1.988&{\it      -0.19}\\
             \allo{H}{I  }{4340}&3& {\bf     47.000}&    47.341&{\it       0.04} &             \allo{N}{II }{4530}&0& {\bf      0.004}&     0.004&{\it      -0.53} &             \forb{S}{II }{6731}&0& {\bf      4.420}&     3.846&{\it      -0.76}\\
             \allo{H}{I  }{6563}&0& {\bf    312.000}&   291.434&{\it      -0.71} &        \allo{N}{II }{4552}$^a$ &0& {\bf      0.003}&     0.001&{\it      -2.79} &             \forb{S}{II }{6731}&3& {\bf      6.790}&     5.174&{\it      -1.04}\\
             \allo{H}{I  }{6563}&3& {\bf    350.000}&   291.063&{\it      -1.01} &        \allo{N}{II }{5495}$^a$ &0& {\bf      0.015}&     0.003&{\it      -5.99} &             \forb{S}{II }{6731}&4& {\bf      4.270}&     4.092&{\it      -0.16}\\
             \allo{H}{I  }{2.62}&1& {\bf      4.718}&     4.855&{\it       0.09} &        \allo{N}{II }{5679}$^a$ &0& {\bf      0.067}&     0.017&{\it      -5.32} &             \forb{S}{III}{6312}&0& {\bf      0.857}&     0.873&{\it       0.07}\\
             \allo{H}{I  }{4.05}&1& {\bf      8.557}&     8.679&{\it       0.04} &        \allo{N}{III}{4641}$^a$ &0& {\bf      0.005}&     0.000&{\it     -38.03} &             \forb{S}{III}{6312}&3& {\bf      0.940}&     0.925&{\it      -0.05}\\
             \allo{H}{I  }{7.45}&1& {\bf      2.731}&     2.815&{\it       0.09} &             \forb{O}{I  }{5577}&0& {\bf      0.026}&     0.017&{\it      -1.59} &             \forb{S}{III}{6312}&4& {\bf      0.960}&     0.865&{\it      -0.31}\\
            \allo{He}{I  }{4026}&0& {\bf      2.098}&     2.005&{\it      -0.25} &             \forb{O}{I  }{6300}&0& {\bf      2.170}&     1.137&{\it      -3.55} &             \forb{S}{III}{9069}&0& {\bf     17.800}&    18.799&{\it       0.57}\\
            \allo{He}{I  }{4471}&0& {\bf      4.490}&     4.280&{\it      -0.26} &             \forb{O}{I  }{6300}&4& {\bf      2.040}&     1.302&{\it      -1.71} &             \forb{S}{III}{9069}&3& {\bf     14.550}&    18.896&{\it       1.43}\\
            \allo{He}{I  }{4471}&3& {\bf      3.590}&     3.288&{\it      -0.34} &             \forb{O}{I  }{6363}&0& {\bf      0.759}&     0.363&{\it      -2.82} &             \forb{S}{III}{9532}&0& {\bf     42.200}&    46.621&{\it       1.04}\\
            \allo{He}{I  }{4471}&4& {\bf      4.170}&     4.116&{\it      -0.05} &             \forb{O}{II }{2471}&2& {\bf     19.271}&    18.374&{\it      -0.18} &             \forb{S}{III}{9532}&3& {\bf     38.090}&    46.861&{\it       1.14}\\
            \allo{He}{I  }{4713}&0& {\bf      0.610}&     0.647&{\it       0.22} &             \forb{O}{II }{3726}&0& {\bf    123.800}&    88.632&{\it      -3.51} &             \forb{S}{III}{18.6}&1& {\bf     15.204}&    16.078&{\it       0.21}\\
            \allo{He}{I  }{5876}&0& {\bf     13.670}&    11.557&{\it      -1.76} &             \forb{O}{II }{3726}&3& {\bf     90.530}&   115.323&{\it       1.33} &             \forb{S}{III}{33.4}&1& {\bf      2.617}&     2.600&{\it      -0.02}\\
            \allo{He}{I  }{5876}&3& {\bf     10.850}&     8.885&{\it      -1.10} &             \forb{O}{II }{3729}&0& {\bf     52.340}&    34.851&{\it      -4.27} &             \forb{S}{IV }{10.5}&1& {\bf      1.266}&     2.929&{\it       2.49}\\
            \allo{He}{I  }{5876}&4& {\bf     10.200}&    11.102&{\it       0.47} &             \forb{O}{II }{3729}&3& {\bf     37.720}&    45.228&{\it       1}&            \forb{Cl}{II }{8579}&0& {\bf      0.284}&     0.183&{\it      -1.69}\\
            \allo{He}{I  }{6678}&0& {\bf      3.870}&     2.960&{\it      -1.47} &             \forb{O}{II }{7323}&0& {\bf     13.740}&    13.333&{\it      -0.32} &            \forb{Cl}{II }{8579}&3& {\bf      0.270}&     0.248&{\it      -0.25}\\
            \allo{He}{I  }{6678}&3& {\bf      2.550}&     2.283&{\it      -0.42} &             \forb{O}{II }{7323}&3& {\bf     15.680}&    18.152&{\it       0.80} &            \forb{Cl}{II }{9124}&0& {\bf      0.077}&     0.048&{\it      -1.78}\\
            \allo{He}{I  }{6678}&4& {\bf      2.750}&     2.845&{\it       0.13} &             \forb{O}{II }{7332}&0& {\bf     11.500}&    10.733&{\it      -0.72} &            \forb{Cl}{II }{9124}&3& {\bf      0.050}&     0.066&{\it       0.80}\\
            \allo{He}{I  }{7065}&0& {\bf      7.040}&     8.266&{\it       0.88} &             \forb{O}{II }{7332}&3& {\bf     13.380}&    14.613&{\it       0.48} &            \forb{Cl}{II }{6162}&0& {\bf      0.004}&     0.004&{\it       0.28}\\
            \allo{He}{I  }{7065}&4& {\bf      5.130}&     7.867&{\it       1.63} &             \allo{O}{I  }{7773}&0& {\bf      0.075}&     0.054&{\it      -1.29} &            \forb{Cl}{III}{5518}&0& {\bf      0.182}&     0.179&{\it      -0.05}\\
            \allo{He}{I  }{7281}&0& {\bf      0.791}&     0.683&{\it      -0.56} &        \allo{O}{I  }{8447}$^a$ &0& {\bf      1.142}&     0.007&{\it     -28.20} &            \forb{Cl}{III}{5518}&3& {\bf      0.190}&     0.170&{\it      -0.33}\\
             \forb{C}{I  }{8727}&0& {\bf      0.033}&     0.030&{\it      -0.39} &             \allo{O}{I  }{9264}&0& {\bf      0.027}&     0.020&{\it      -1.06} &            \forb{Cl}{III}{5518}&4& {\bf      0.450}&     0.182&{\it      -2.69}\\
             \forb{C}{II }{157.}&1& {\bf      1.100}&     0.403&{\it      -2.98} &             \forb{O}{III}{4363}&0& {\bf      0.935}&     0.710&{\it      -1.05} &            \forb{Cl}{III}{5538}&0& {\bf      0.356}&     0.356&{\it       0.00}\\
            \sforb{C}{II }{2326}&2& {\bf     81.462}&    88.482&{\it       0.31} &             \forb{O}{III}{4363}&3& {\bf      0.520}&     0.421&{\it      -0.63} &            \forb{Cl}{III}{5538}&3& {\bf      0.380}&     0.348&{\it      -0.26}\\
            \sforb{C}{III}{1909}&2& {\bf     27.592}&    34.878&{\it       0.89} &             \forb{O}{III}{4363}&4& {\bf      0.910}&     0.716&{\it      -0.71} &            \forb{Cl}{III}{5538}&4& {\bf      1.000}&     0.353&{\it      -3.09}\\
             \allo{C}{II }{1335}&2& {\bf     23.212}&    20.164&{\it      -0.54} &             \forb{O}{III}{4959}&0& {\bf     72.700}&    65.158&{\it      -1.15} &            \forb{Cl}{III}{8436}&0& {\bf      0.006}&     0.010&{\it       1.68}\\
        \allo{C}{II }{1761}$^a$ &2& {\bf      2.102}&     0.349&{\it      -5.34} &             \forb{O}{III}{4959}&3& {\bf     29.520}&    37.300&{\it       1.28} &            \forb{Cl}{III}{8483}&0& {\bf      0.011}&     0.010&{\it      -0.29}\\
             \allo{C}{II }{4267}&0& {\bf      0.570}&     0.452&{\it      -0.88} &             \forb{O}{III}{4959}&4& {\bf     50.100}&    65.034&{\it       1.43} &            \forb{Cl}{III}{8504}&0& {\bf      0.004}&     0.009&{\it       2.80}\\
             \allo{C}{II }{4267}&4& {\bf      0.690}&     0.436&{\it      -1.36} &             \forb{O}{III}{5007}&0& {\bf    214.000}&   196.126&{\it      -0.92} &            \forb{Ar}{II }{6.98}&1& {\bf      5.997}&     6.184&{\it       0.09}\\
             \allo{C}{II }{4619}&0& {\bf      0.011}&     0.021&{\it       2.61} &             \forb{O}{III}{5007}&3& {\bf     85.870}&   112.273&{\it       1.47} &            \forb{Ar}{III}{5192}&0& {\bf      0.039}&     0.034&{\it      -0.43}\\
        \allo{C}{II }{6580}$^a$ &0& {\bf      0.805}&     0.039&{\it     -11.52} &             \forb{O}{III}{5007}&4& {\bf    151.000}&   195.754&{\it       1.42} &            \forb{Ar}{III}{5192}&3& {\bf      0.030}&     0.027&{\it      -0.28}\\
        \allo{C}{II }{7231}$^a$ &0& {\bf      0.169}&     0.006&{\it     -12.66} &             \forb{O}{III}{51.8}&1& {\bf     15.261}&    13.430&{\it      -0.49} &            \forb{Ar}{III}{7135}&0& {\bf      8.260}&     7.736&{\it      -0.36}\\
             \forb{N}{I  }{5198}&0& {\bf      0.201}&     0.046&{\it      -5.58} &             \forb{O}{III}{88.3}&1& {\bf      2.407}&     1.854&{\it      -0.77} &            \forb{Ar}{III}{7135}&3& {\bf      6.180}&     5.937&{\it      -0.15}\\
             \forb{N}{I  }{5200}&0& {\bf      0.117}&     0.016&{\it      -7.69} &             \allo{O}{II }{4152}&0& {\bf      0.018}&     0.023&{\it       0.88} &            \forb{Ar}{III}{7135}&4& {\bf      6.030}&     7.417&{\it       0.79}\\
             \forb{N}{II }{5755}&0& {\bf      2.760}&     2.599&{\it      -0.33} &             \allo{O}{II }{4341}&0& {\bf      0.085}&     0.091&{\it       0.23} &            \forb{Ar}{III}{7751}&0& {\bf      2.196}&     1.867&{\it      -0.89}\\
             \forb{N}{II }{5755}&3& {\bf      3.910}&     3.723&{\it      -0.19} &        \allo{O}{II }{4593}$^a$ &0& {\bf      0.024}&     0.004&{\it      -6.37} &            \forb{Ar}{III}{7751}&3& {\bf      0.780}&     1.433&{\it       1.81}\\
             \forb{N}{II }{5755}&4& {\bf      2.190}&     2.694&{\it       0.79} &             \allo{O}{II }{4651}&0& {\bf      0.171}&     0.172&{\it       0.01} &            \forb{Ar}{III}{9.00}&1& {\bf      7.334}&     8.039&{\it       0.27}\\
             \forb{N}{II }{6548}&0& {\bf     53.600}&    53.094&{\it      -0.10} &            \forb{Ne}{II }{12.8}&1& {\bf     53.098}&    61.004&{\it       0.53} &            \forb{Ar}{IV }{4711}&0& {\bf      0.003}&     0.003&{\it       0.08}\\
             \forb{N}{II }{6548}&3& {\bf     71.230}&    73.639&{\it       0.18} &            \forb{Ne}{III}{3869}&0& {\bf      3.090}&     2.806&{\it      -0.53} &            \forb{Ar}{IV }{4740}&0& {\bf      0.004}&     0.005&{\it       1.58}\\
             \forb{N}{II }{6548}&4& {\bf     51.300}&    54.902&{\it       0.37} &            \forb{Ne}{III}{3869}&3& {\bf      2.050}&     1.147&{\it      -2.21} &            \forb{Fe}{III}{4659}&0& {\bf      0.027}&     0.027&{\it      -0.06}\\
             \forb{N}{II }{6584}&0& {\bf    162.900}&   156.681&{\it      -0.41} &            \forb{Ne}{III}{3869}&4& {\bf      2.630}&     3.766&{\it       1.37} &            \forb{Fe}{III}{5271}&0& {\bf      0.015}&     0.017&{\it       0.34}\\
             \forb{N}{II }{6584}&3& {\bf    206.830}&   217.309&{\it       0.27} &            \forb{Ne}{III}{3968}&0& {\bf      0.970}&     0.846&{\it      -0.52} &            \forb{Fe}{III}{4881}&0& {\bf      0.015}&     0.013&{\it      -0.52}\\
             \forb{N}{II }{6584}&4& {\bf    151.000}&   162.014&{\it       0.39} &            \forb{Ne}{III}{3968}&3& {\bf      0.640}&     0.346&{\it      -1.83} &            \allo{Mg}{II }{2798}&2& {\bf     16.351}&    15.697&{\it      -0.16}\\
             \forb{N}{II }{121.}&1& {\bf      0.283}&     0.189&{\it      -1}&            \forb{Ne}{III}{15.5}&1& {\bf      9.512}&     9.276&{\it      -0.08} &            \forb{Si}{II }{34.8}&1& {\bf      0.936}&     0.985&{\it       0.13}\\

    \hline
  \end{tabular}}
{\footnotesize $^a$ Lines suspected to be mainly emitted by fluorescence.}
\end{table*}

\begin{figure}
  \centering
    \includegraphics[width=8.cm]{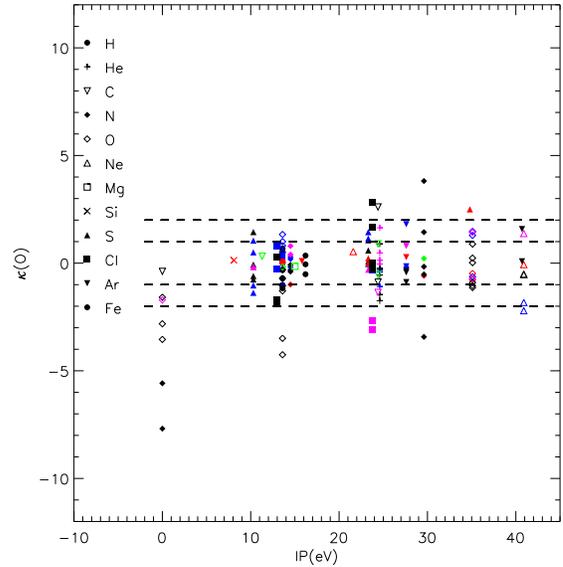}
  \caption{Quality factors $\kappa({\rm O})$ vs. the ionization potential of the emitting ion. Each symbol corresponds to a given element.
  }
  \label{fig:QF.IP}
\end{figure}

Table~\ref{tab:res1} presents the intensity ratios obtained for our best-fit model and compares them with the observed values. The first column is the identification of the emission line. Some lines appear more than once because they have been observed by various authors and through different apertures. The second column provides the reference of the observation, using the following code: 0 for CTIO \citep{2004ApJ...615..323S}; 1 for ISO \citep{2004A&A...423..593P}; 2 for IUE \citep{2004A&A...423..593P}; 3 for Lick \citep{1994PASP..106..745H}; 4 for CTIO \citep{1985PASP...97..397G}.
The third and fourth columns are the observed values and the results of the model, respectively, in unit H$\beta$ = 100. The fifth column is $\kappa({\rm O})$, the quality factor of the fit, as defined in Sect.~\ref{sec:definning-tolerance}. 
Figure \ref{fig:QF.IP} shows all the values of $\kappa({\rm O})$ versus the ionization potentials (IPs) of the ions producing the lines. Horizontal lines at $\kappa({\rm O}) = \pm 1, \pm 2$ help us to see the lines that are fitted within acceptable tolerance limits.

\begin{figure}
  \centering
    \includegraphics[width=8.cm]{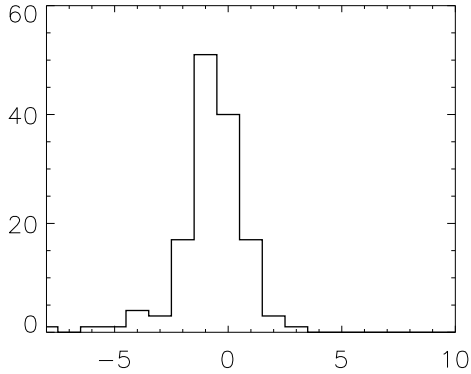}
  \caption{Histogram of $\kappa({\rm O})$ distribution.}
  \label{fig:histo}
\end{figure}

Figure \ref{fig:histo} shows the distribution of the values of $\kappa({\rm O})$. The symbols are related to the emitting element. Most of the values of $\kappa({\rm O})$ are in the [-1,1] range, that is, the value obtained by the model agrees with the observed one within the tolerance limits. 

We do not observe a stronger discrepancy between the model predictions and the observed intensities at lower intensities. We interpret this as an indication that the tolerances defined in Sect.\ref{sec:definning-tolerance} are adequate.

The hydrogen recombination lines are almost all reproduced well, with the notable exception of the ratio H$\alpha/$H$\beta$, for which the observed values are 3.5 and 3.1 (depending on the observer), considerably higher than the predicted value of 2.9. We can suspect this to be an observational problem.

The helium recombination lines are also mostly fitted within the tolerance levels. When a given line is not fitted (e.g., \alloa{He}{i}{5876}), this is true for only one observation of the given line, the other observations of the same line are reproduced well ($\kappa(\alloa{He}{i}{5876}) = -1.74, -1.09, 0.47$ depending on the observation). 

The 3 most intense lines of carbon are reproduced within the tolerance levels. 

The \dforba{N}{ii}{6548}{6584} lines are reported by 3 authors, and the corresponding 6 observed intensities are reproduced well with a mean $\kappa(\dforba{N}{ii}{6548}{6584})$ of 0.13. We note the very small value predicted for \forba{N}{i}{5200}, 8 times smaller than observed. A discussion of the electron excitation collision strengths of N$^+$ can be found in \citet{2001ASPC..247..533P}, who questioned the values published by \citet{2000ADNDT..76..191T}, but the version of Cloudy that we used (namely c07.02.01) assumes the collision strengths of \citet{2006ApJS..163..207T}, which are supposed to be 20 times greater at an electron temperature of about 8000~K. This line is emitted by the very external part of the nebula. Any small variation in the position of the slit or the external electron density could strongly affect the prediction.

The \ion{O}{i} lines are globally predicted to be half of the observed values. They are also emitted close to the edge of the nebula, and the same considerations as for [\ion{N}{i}] lines about the slit position relative to the main sites of emission should apply. The [\ion{O}{ii}] lines are reproduced well, with the exception of the \dforba{O}{ii}{3726}{3729} observed by \citet{2004ApJ...615..323S}, which are observed 40\% higher than the model value. On the other hand, the observed intensities of the same lines reported by \citet{1994PASP..106..745H} are 25\% lower than the model values. The \dforba{O}{ii}{7320}{7330} lines are better reproduced, with the same tendencies. The \dforba{O}{iii}{4959}{5007} lines are reported by 3 authors, the corresponding 6 intensities are reproduced with a mean $\kappa(\dforba{O}{iii}{4959}{5007})$ of 0.6. The observed value for \forba{O}{iii}{5007}/\hbeta/ ranges from 0.86 to 2.1 depending on the author. The values predicted by the model range from 1.1 to 2.0, clearly showing the effect of changing the position and size of the aperture and the need to take both into account in the fitting process.

The 5 values reported in the literature for the [\ion{Ne}{iii}] lines are reproduced with a mean value for $\kappa([\ion{Ne}{iii}])$ of -0.6, while $\kappa([\ion{Ne}{ii}]12.8 \mu\rm{m}) \sim 0.5$.

The 10 values reported for the [\ion{S}{ii}] lines are reproduced with a mean $\kappa([\ion{S}{ii}])$ of -0.12, while we found a mean $\kappa([\ion{S}{iii}])$ of 0.46, using 9 lines of [\ion{S}{iii}]. The mean value of $\kappa([\ion{Cl}{ii}])$ is -0.52 (5 lines) and the mean value of $\kappa([\ion{Cl}{iii}])$ is -0.24 (9 lines). For the Ar lines, the mean value for $\kappa([\ion{Ar}{iii}])$ is 0.10, based on 8 lines. 

All of these mean values for $\kappa(\rm{O})$ are only indicative, as some very faint lines can enter into the mean and dominate it. Figure~\ref{fig:QF.IP} is very instructive for obtaining an idea of the global quality of the fit. Notice that for Mg and Si only one line of each element is detected.

Some lines are permitted lines that are suspected of having a strong fluorescence contribution, and for which $\kappa({\rm O})$ is much lower than -1 \citep[see][ for the NII fluorescence spectrum]{2005MNRAS.361..813E}. They are labeled by an $^a$ close to the wavelength. For example, the \alloa{N}{iii}{4641} line is virtually not produced by recombination of the N$^{+++}$ ion (requiring 47~eV photons), but rather is emitted entirely by fluorescence, implying that the N$^{++}$ ion is responsible for the emission.

\begin{figure*}
  \centering
    \includegraphics[width=15.cm]{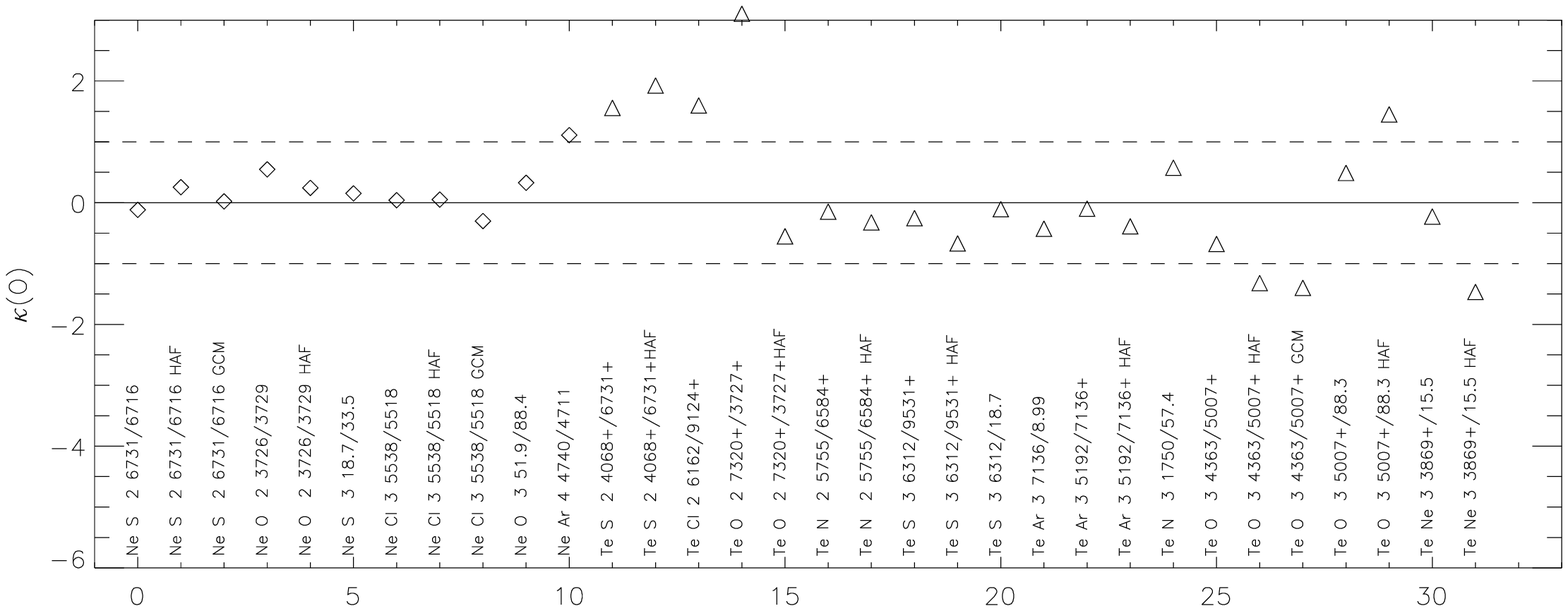}
  
  \caption{Quality factors $\kappa({\rm O})$ for the diagnostic ratios (diamonds for electron density and triangles for electron temperature) from Tab.~\ref{tab:diag1}. For the optical lines, HAF refers to \citet{1994PASP..106..745H}, GCM to \citet{1985PASP...97..397G}, and otherwise to \citet{2003ApJS..149..157S}. Other observations are described in Sect.~\ref{sec:nebular-observations}.
  }
  \label{fig:resdiag}
\end{figure*}

Table~\ref{tab:diag1} (online only) compares the observed electron density and temperature diagnostics with the results of the model.  The four columns correspond to I: diagnostic line ratio, II: observed ratio, III : model ratio, IV:  quality factor $\kappa({\rm O})$ (see Sect.\ref{sec:definning-tolerance}) for the model. All the density (temperature) diagnostic ratios increase with the electron density (temperature resp.). For the optical lines, HAF refers to \citet{1994PASP..106..745H}, GCM to \citet{1985PASP...97..397G}, and otherwise to \citet{2003ApJS..149..157S}. These results are also plotted in Fig.~\ref{fig:resdiag}.
From the 11 density diagnostics that are available from the observed line ratios, we find very good agreement between the model and the observed values. The only value of $\kappa({\rm O})$ that is not in the [-1,1] range is \rforba{Ar}{iv}{4740}{4711}, perhaps indicative of a slightly higher density thin shell or clumps close to the star. The mean value of $\kappa({\rm density diagnostics})$, without taking into account the \rforba{Ar}{iv}{4740}{4711} ratio, is 0.12.

Twenty-one diagnostic line ratios are available for the electron temperature, and for most of them the observed ratios agree with the results of the model. We nevertheless note the disagreement for the low ionization species (\ion{S}{ii} and \ion{Cl}{ii}). The extreme value of 3.1 is obtained for $\kappa(T_e(\rforba{O}{ii}{7320}{3727}))$ for the \citet{2003ApJS..149..157S} observation. 
On the other hand, for the same diagnostic ratio using \citet{1994PASP..106..745H} observations, we find $\kappa({\rm O}) = -0.2$. This demonstrate the difficulties of the fitting exercise, especially when relying on a unique set of observations. The same applies to the classical $T_e(\rforba{O}{iii}{5007}{4363})$ diagnostic, which is rather well reproduced when comparing to the \citet{2003ApJS..149..157S} observations, but less well fitted when using the \citet{1994PASP..106..745H} or \citet{1985PASP...97..397G} observations. For most of the diagnostics for which we have $|\kappa({\rm O})| > 1$, there is another observation of the same diagnostic for which $|\kappa({\rm O})| < 1$ (see Sect.~\ref{sec:power} for a discussion of the small discrepancy between the model and the observed electron temperature diagnostics).

Figure~\ref{fig:reslambda} (online only) shows that there is no systematic trend for $\kappa({\rm O})$  as a function of either the wavelength (upper panel) or the critical density of the forbidden lines (lower panel). The first result indicates that there is no important error in the reddening correction applied to the emission line intensities. If there is a low density inter-clump medium, it could be detected by low critical density lines. The absence of a clear trend in the lower panel of Fig.~\ref{fig:reslambda} indicates that there is no detectable low density medium: this could mean that there are no clumps at all and the medium is rather homogeneous, or that any inter-clump medium is of a very low density (see Sects.~\ref{sec:clumping-factor-and-D} and \ref{sec:evol-tracks} for a discussion of the determination of the clumping factor).

\label{sec:absolute-fluxes}
Table~\ref{tab:absol} displays the predicted absolute fluxes in the 4 lines used to cross-calibrate different spectral regimes, and the corresponding observed values. The agreement is extremely good. In this way we verify that our cross-calibration between the different wavelength regions, described in Sect.~\ref{sec:nebular-observations} , is good.
 
\begin{table}\centering
  \caption{Comparison of some predicted and observed absolute fluxes (units in $10^{-10}$ erg.cm$^{-2}$.s$^{-1}$, except for the 6cm continuum, in Jy). The model \hbeta/ and \forba{O}{ii}{2471} fluxes are reddened using E$_{\rm B-V} =$\cmebv\ and R$_{\rm V} = $ \cmrv. The observed (reddened) values are taken from \citet{2004A&A...423..593P}} 
  \label{tab:absol}
  \begin{tabular}{|l|r|r|}
    \hline \noalign{\smallskip}
Line & Observation & Model \\
\noalign{\smallskip} \hline \noalign{\smallskip}
\hbeta/  & 2.7 &  2.56 \\
\forba{O}{ii}{2471} & 0.24 &  0.21 \\
\allom{H}{i}{4.05} & .45 &  0.59 \\
6 cm continuum & 1.73 &  1.96 \\
    \noalign{\smallskip} \hline
  \end{tabular}
\end{table}

             \subsection{Fitting the IUE observations}
\label{sec:iue}

\begin{figure}
  \centering
    \includegraphics[width=9.cm]{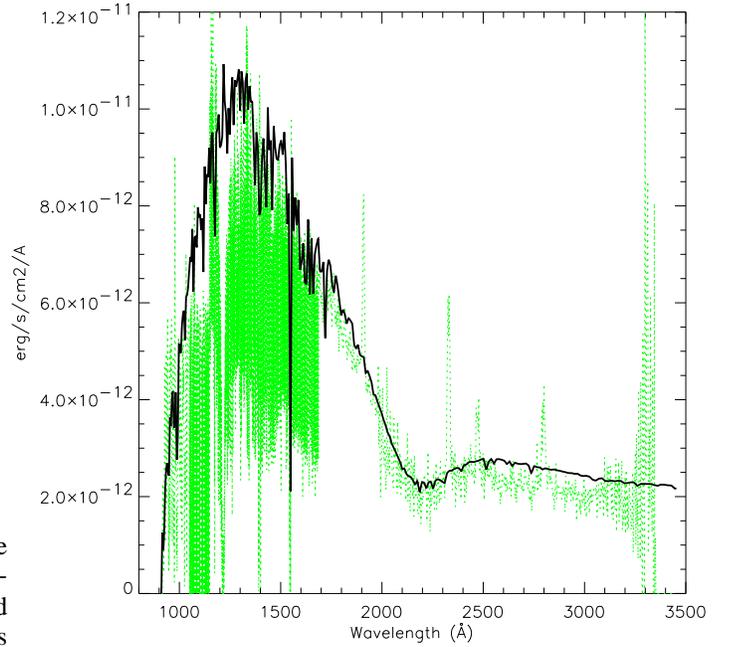}  
  \caption{Comparison between the IUE observations and the stellar SED from the CMFGEN model.}
  \label{fig:compIUE}
\end{figure}

Figure~\ref{fig:compIUE} compares the observed IUE stellar spectrum with the stellar model used in the photoionization model. The model shown here is the CMFGEN star taking into account the nebular absorption and the ISM attenuation. We found that the best-fit extinction law to reproduce the IUE observations is obtained using the \citet{1999PASP..111...63F} law with E$_{\rm B-V} =$\cmebv\ and R$_{\rm V} = $ \cmrv. The 2200\AA\ bump is reproduced well, as well as the EUV ISM attenuation. This value for E$_{\rm B-V} $ is within the range obtained from the literature \citep[from 0.14 to 0.37, see][and references therein]{2004A&A...423..593P,2004ApJ...615..323S}.

             \section{Discussion}
\label{sec:discussion}

             \subsection{Missing power}
\label{sec:power}

We point out that, in contrast to the electron density, which is an input parameter of the model (being mainly controlled by the hydrogen density), the electron temperature is an output parameter, as it is the result of the energy balance in the nebula, and depends on the ionizing spectrum and the metal abundances. Thus the electron temperature diagnostics are predictions for the present model, because all the parameters that can have an impact on the temperature are fixed: the ionizing SED is constrained by the stellar observations and the abundances of the major coolants are controlled by emission lines that are already fitted. Even if the fits to the electron temperature diagnostic line is relatively good,  the model presented here seems to be a little too cool. On the other hand, the fit to the intensities of the collisionally excited lines is also globally quite low (see the distribution of $\kappa({\rm O})$ in Fig.\ref{fig:histo}). This small discrepancy between the model and observations could be resolved by adding an extra heating process.

We test a model with an extra heating value of 10$^{-20.3} \times (n_H / n_0) $~erg.cm$^{-3}$.s$^{-1}$ (for this we used c08.00 version of Cloudy, with $n_0=1$~cm$^{-3}$), to obtain electron temperature diagnostics closer to the observations. This leads to an increase in the mean electron temperature from 8575~K to 9046~K. The corresponding total additional power is 4 $\times 10^{35}$~erg.s$^{-1}$, representing 30\% (!) of the heating from the star (or the cooling of the nebula, since they are equal by assumption of thermal equilibrium). The collisionally excited lines in this boosted model are globally closer to the observed values. 

             \subsection{Where a black-body model fails}
\label{sec:bb}

To test the validity of using a simple SED for the ionizing flux instead of an atmosphere model, we perform a model with a black-body (BB) distribution. We keep all model parameters to the values of the model described above, except for the stellar temperature, which is set to be \cmbbtemp~kK to produce a global ionization of the nebula similar to what is obtained with the CMFGEN model.
Figures~\ref{fig:QF.IP-BB} and~\ref{fig:histo-BB} (online only) are similar to Figs.~\ref{fig:QF.IP} and~\ref{fig:histo} respectively, except that they involve this BB model.

\begin{figure*}
  \centering
    \includegraphics[width=16.cm]{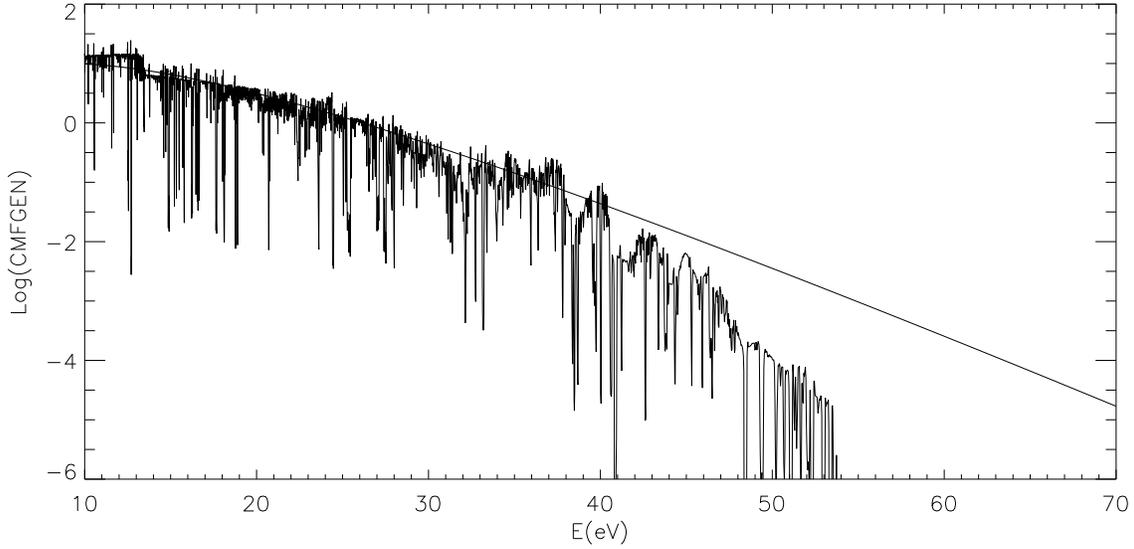}
  
  \caption{This is the comparison between the CMFGEN model and a black body at \cmbbtemp~kK.}
  \label{fig:compio}
\end{figure*}

All the elements apart from Ne and Ar exhibit line intensities very similar to those obtained with the adopted model. For Ar and Ne, the ions of high IPs $>$ 40~eV (namely Ne$^{2+}$ and Ar$^{3+}$) are strongly overpredicted by the BB model, relative to the lower IPs ions from the same species. We attribute these overpredictions to the shape of the SED at high energy (see Fig.\ref{fig:compio}). The absorption features visible between 40 and 50~eV are mainly produced by \ion{C}{iii}, \ion{O}{iv}, and \ion{Ne}{iii}, among other lines.

The main difference between the BB model and the CMFGEN model is only visible above 40eV. For energies below this limit, the two SEDs are very similar. Thus, in the absence of any atmosphere model, the Planck hypothesis is a good approximation to the ionizing flux between 10 and 40~eV, but there is a significant  shift in the effective temperature (here a Planck function at \cmbbtemp~kK is similar to an atmosphere model close to \cmcmftemp~kK).

We note that \citet{1994PASP..106..745H} obtained very low abundances for Ne and Ar using a photoionization model of IC418. We suspect that the SED that they used is responsible for this discrepancy with our results. 

             \subsection{Clumping factor and distance determination}
\label{sec:clumping-factor-and-D}

None of the geometrical size of the nebula, the distance, nor the absolute luminosity of the central star are well-constrained. Multiplying the stellar luminosity by a factor of $K_1$ but instead all the sizes of the nebula (inner radius, positions and sizes of the 2 shells)  by $K_1^{1/3}$ will lead to the same position of the recombination front and the same appearance (because the Str\"omgren radius is proportional to Q$_0^{1/3}$). If the distance to the observer is also multiplied by $K_1^{1/3}$, the angular size remains constant, while the absolute luminosity (\hbeta/ for example) is multiplied by the same factor $K_1^{1/3}$. On the other hand, changing Q$_0$ and the sizes by factors of $K_1$ and $K_1^{1/3}$, respectively, leads to increase in the ionization parameter of $U = Q_0/(4.\pi.R^2.N_H.c) \propto K_1^{1/3}$. This will lead to an increase in the intensities of the lines emitted by high IP ions (e.g., \ion{O}{iii}, \ion{Ar}{iv}, \ion{Ne}{iii}) and a decrease in the intensities of the low IP lines (e.g., \ion{S}{ii}). This behaviour is very similar to the reaction of the nebula when the stellar temperature increases. 
We use the \hbeta/ absolute flux to resolve the degeneracy and to fine-tune the effective temperature of the star (to within an uncertainty level of 500~K, the first order value being obtained from the stellar spectrum fit) by reproducing the ionization state of the nebula.

Another factor that can strongly increase the uncertainty in the morphology (and distance) determination is the filling factor.  This factor can be divided into 2 parts: a morphological filling factor $ff_M$ and a clumping factor $ff_C$.
The first one represents the ratio of the volume of the gaseous nebula to the volume of the total extension of the nebula. In the simple case of a thin shell of external radius $R_{neb}$ surrounding an empty cavity of radius $R_{cav}$ , we have $ff_M = 1-(\frac{R_{cav}}{R_{neb}})^3$. The use of a detailed description of the nebular density structure fixes this morphological filling factor. On the other hand, the clumping filling factor aims to take into account any possible small-scale structure of the nebula, where the gas would be concentrated into small clumps. The physical process that could lead to this situation is unclear; the real situation is certainly not clumps in the middle of a vacuum, but this filling factor is often used as the first moment of a more complex distribution. This is the goal of the ``filling factor'' keyword used in Cloudy. The global action of this factor is to artificially increase the geometrical size of the nebula. There is virtually no difference between observables of a model with $Q_0$, $S$ and $ff_C$ and a model with $K_2^2\times Q_0$, $K_2\times S$ and $ff_C /K_2$, where $Q_0$ is the number of ionizing photons emitted by the central source, $S$ is a parameter describing all the sizes of the nebula ($R_1$ and  $R_2$ for example in Eq. \ref{eq:morf}), and $K_2$ is a coefficient greater than 1. All the line intensities (absolute and relative), the angular sizes, and the images  are exactly the same in both models. The nebula would be at a distance $D$ in the first case and $K_2\times D$ in the second. Once the existence of this clumping factor is accepted, then there is no way of determining the distance to the nebula, nor the absolute luminosity of the star, in contrast to the claims of either \citet{2005ApJ...620..321M} or \citet{2006A&A...451..925G}.

Using a filling factor of 1.0 as a first guess (i.e. no clumps), we determine the following parameters:
the \hbeta/ flux value (without reddening) is predicted to be $7.2\times 10^{-10}$ erg cm$^{-2}$ s$^{-1}$.
The distance is estimated to be $K_2\times 1.26$~kpc.
The ionized hydrogen mass is $K_2^2\times 0.057$ solar masses.
With an equatorial radius of $K_2\times1.1\times 10^{17}$cm and considering an expansion velocity of 30 km/s (see Sect.~\ref{sec:line-profiles}), the age of the nebula is $K_2\times \cmage$ years.
The stellar luminosity is $K_2^2\times \cmlumi$ solar luminosities and the corresponding number of ionizing photons is $K_2^2\times 10^{47.6} \rm{s}^{-1}$. 

Because of the presence of stellar wind, it is impossible to obtain the absolute luminosity of the star and so likewise impossible to determine its spectroscopic distance.

The distance to IC418 was determined by \citet{2009AJ....138...46G}, using parallax expansion between two VLA observations with more than 20 years of time delay. They determined a distance of 1.3 $\pm 0.4$kpc. This value leads to a determination of the parameter $K_2$ close to unity and the other $K_2$-dependant parameters as given in Table~\ref{tab:param}.

In the following section, we present an independent (and more indirect) way of determining the distance to the PN.
             \subsection{Using evolutionary tracks to determine the distance}
\label{sec:evol-tracks}
\label{sec:distance}

Once the effective temperature is known and using the evolutionary tracks from \citet{1994ApJS...92..125V} or \citet{1995A&A...299..755B}, a simple relation between the stellar luminosity and the evolutionary age of the nebula can be obtained. For \teff/ close to \cmcmftemp~kK, the relation is log(age)$ = 17.34-3.73\times \rm{log}(L/L_{\odot}$) from \citet{1994ApJS...92..125V} and log(age)$ = 20.6-4.53 \times \rm{log}(L/L_{\odot}$) from \citet{1995A&A...299..755B}. 

On the other hand, from Sec.\ref{sec:clumping-factor-and-D} we determined the age to be $K_2\times \cmage$ years and the luminosity to be $K_2^2\times \cmlumi$ solar luminosities. We then have: age(years)$ = \cmage \times \sqrt{L/\cmlumi}$, similar to log(age)$ = \cmcoeff + 0.5 \times \rm{log}(L/L_{\odot}$).

The intersection of this relation with those determined from evolutionary tracks lead to a determination of the age, the luminosity, and all the parameters depending on $K_2$. A value of $K_2=0.92$ (0.97) is found when using \citet{1994ApJS...92..125V} \citep[][respectively]{1995A&A...299..755B}, leading to a clumping factor very close to 1.

The results presented in this section strongly depend on the evolutionary tracks on the one hand, but on the other hand also on the determination of the dynamical age of the nebula. This latest is supposed to be the ratio of the size of the nebula to the expansion velocity. This implicitly assumes that the expansion velocity is constant during the expansion of the nebula, which is the simplest hypothesis. Observations by \citet{2008ApJ...689..203R} seems to indicate that the expansion is accelerating, confirming theoretical predictions on the RAM pressure acceleration of the AGB envelope \citep[][and references therein]{2006ApJ...646L..61G}. In this case, the age that we determine from the line profiles is only an upper limit to the true dynamical age of the nebula, and this will induce a shift in the position of IC418 in the evolutionary tracks. For example, if we consider that the expansion velocity is only half the value that we used in the previous section, to take into account the acceleration, the value obtained for $K_2$ is reduced by 0.03 dex. We note that this decrease in the mean expansion velocity leads to a value of $K_2$ lower than 1, and the corresponding filling factor will be $1/K_2 > 1$.
 
The method presented in this section for determining a distance from evolutionary tracks clearly depends on the reliability of these latest. The results are comparable when the tracks from either \citet{1994ApJS...92..125V} or \citet{1995A&A...299..755B} are used. The derived filling factor is close to unity, and the distance is comparable to the one obtained by the parallax method by \citet{2009AJ....138...46G}, which all implies that the evolutionary tracks are trustable, at least in the part of the HR-diagram occupied by IC418.

             \subsection{Comparing the stellar and the nebular abundances}
\label{sec:comp-stell-nebul}

The nebula is found to be both carbon- and nitrogen-rich (0.5 and 0.2 dex above the solar value, respectively) while the oxygen abundance is close to solar (0.05 dex lower). 
The neon and argon abundances are close to the solar values. Magnesium, silicon, sulfur and chlorine are underabundant by 0.5 dex relative to their solar values, while iron is deficient by 2.9 dex, suggesting the depletion of those elements in dust grains (whose presence is supported by the IR continuum emission). The nebular helium abundance is 0.2 dex greater than its solar value.

The star is also found to be carbon-rich. The values obtained for the ratio C/O (by mass) are comparable: 1. and 1.5 for the star and the nebula respectively. The uncertainty in C/H is quite large in the case of the nebular value. If we only consider the recombination lines of C, we would obtain higher value for C/H, while the forbidden lines would imply a lower value. A closer agreement between the stellar and nebular values of C/O is then achieved when using only the forbidden lines.

We developed a test model by adopting the same abundance for the star and the nebula (equal to the mean value of the two abundances), and the model was unable to fit the observations. This implies that the differences between the star and nebula are greater than the tolerance level we could adopt in the fit of the observations.

We also highlight that the presence of the X-ray component (Sect.\ref{sec:xray}) is not taken into account in the stellar model process. The true ionization state of the irradiated atmosphere can be higher than the one predicted by the model. The effect on the abundances determined for the star is hard to estimate, but would be in the sense of underestimating of the true abundances.

The nebular composition should reflect that of the star during the epoch it was ejected. Therefore, the observed differences in the compositions should reflect the evolution of the star after the ejection. First of all, the He abundance in the star is higher than that of the nebula. One possible reason for this is the stellar wind. The star lose part of its hydrogen envelope increasing its relative abundance of helium. Carbon is enriched in both the star and the nebula, which means that the nebula was ejected after the third dredge-up (TDU) when the carbon-rich material is exposed at the surface. The nitrogen abundance is slightly increased but N/O $\sim$ 0.1. This low value, together with the normal oxygen composition means that no hot bottom burning (HBB) occurred \citep{2009ApJ...690.1130K}. The small increase of N can be attributed to TDU and mixing from a incomplete H burning shell \citep{2009ApJ...696..797C}. Thus, CNO abundances are in accordance with the progenitor mass being less than 4.0 (the limit for HBB). In the previous section, we calculated the age of the nebula and the luminosity of the star from evolutionary tracks. The solution obtained implies progenitor masses of 1.75 M$_\odot$ and 3.0 M$_\odot$ 
calculated with the tracks of \citet{1994ApJS...92..125V} and \citet{1995A&A...299..755B}, respectively.
Thus, these progenitor masses are in agreement with those predicted for models that reproduce the observed CNO abundances.

             \section{Conclusion}
\label{sec:conclusion}

We have presented a combined stellar and nebular model of the planetary nebula IC418 that reproduces all the observational material available: optical and UV stellar observations (lines and continuum) and IR, and deep optical and UV emission-line observations as well as HST images for the nebula. A total of 140 emission lines are considered, most of them being predicted entirely by the model (i.e., are not used for the convergence process). 

The use of a 3D photoionization model for the nebula is not essential for describing the morphology, since IC418 is close to spherical, but it is of primary importance when accounting for the effect of the position, size, and orientation of the apertures used in the different observations. The apertures are used as masks over the emission-line surface brightness maps to compute the line intensities.
The apparent discrepancies between the intensities of a given line reported by different authors are sometimes reproduced well when these slit effects are taken into account.

The agreement between the stellar atmosphere model performed using the CMFGEN code and the photoionization models obtained by Cloudy\_3D is impressive: the high ionization potential ions (Ne$^{2+}$ and Ar$^{3+}$), whose lines are overpredicted when using a simple Planck function as the SED, are perfectly reproduced when a stellar atmosphere model is used. This reflects the effect of line blanketing in the $>$ 40~eV region, which drastically reduces the ionic fraction of these ions.

The possible presence of clumps in the nebula leads to a degeneracy between the distance, the luminosity, and the clumping factor. It is a priori impossible to resolve this discrepancy in the case of IC418 because no spectroscopic distance can be obtained because of another degeneracy between the stellar luminosity and the mass-loss rate. Nevertheless, using the distance determined by the parallax method by \citet{2009AJ....138...46G}, we can resolve the degeneracy and conclude that there are no clumps in the nebula. 

We found that evolutionary tracks allows us to complete another more indirect determination of the distance, close to 1.25~kpc and the age of 1400 years. This distance is in very good agreement with the one obtained using the parallax method.

High spectral resolution nebular emission-line profiles are reproduced by a velocity expansion law that strongly increases with the distance from the center of the nebula.

Reliable abundances are determined for the nebula and the star. We found the nebula to be carbon- and nitrogen-rich while the oxygen abundance is close to solar, as are those of neon and argon. Magnesium, silicon and iron are highly underabundant relative to their solar values, which implies that there has been some depletion of those elements into dust grains. Sulfur and chlorine are also deficient. The nebular helium abundance is 0.2 dex higher than the solar value.
The star is also found to be carbon-rich.

\acknowledgements

We first of all wish to thank Vladimir Escalante who asked CM four years ago for a ``quick'' model of IC418... Here we are!
We thank Juan Echevarria and Rafael Costero for providing optical stellar spectrum of IC418.
We are very grateful to Grazyna Stasi\'nska, Michael Richer, Svetozar Zheokov, Jorge Garcia Rojas and Luc Binette for useful discussions and a careful reading of the manuscript.
We also wish to thank Elena Jimenez Bailon for her help on the analysis of the Chandra observations.
The computations were carried out on a AMD-64bit computer financed by grants PAPIIT IX125304 from DGAPA (UNAM,Mexico).
This work is partly 
supported by grants  PAPIIT IN123309 from DGAPA (UNAM,Mexico), CONACyT-48737, CONACyT-49749, and CONACyT-60967 (Mexico).


\begin{thebibliography}{47}
\expandafter\ifx\csname natexlab\endcsname\relax\def\natexlab#1{#1}\fi

\bibitem[{{Aller} \& {Czyzak}(1983)}]{1983ApJS...51..211A}
{Aller}, L.~H. \& {Czyzak}, S.~J. 1983, \apjs, 51, 211

\bibitem[{{Aller} {et~al.}(1980){Aller}, {Keyes}, {Ross}, \&
  {Czyzak}}]{1980Ap&SS..67..349A}
{Aller}, L.~H., {Keyes}, C.~D., {Ross}, J.~E., \& {Czyzak}, S.~J. 1980, \apss,
  67, 349

\bibitem[{{Asplund} {et~al.}(2005){Asplund}, {Grevesse}, \&
  {Sauval}}]{2005ASPC..336...25A}
{Asplund}, M., {Grevesse}, N., \& {Sauval}, A.~J. 2005, in Astronomical Society
  of the Pacific Conference Series, Vol. 336, Cosmic Abundances as Records of
  Stellar Evolution and Nucleosynthesis, ed. T.~G. {Barnes}, III \& F.~N.
  {Bash}, 25--+

\bibitem[{{Beck} {et~al.}(1981){Beck}, {Lacy}, {Townes}, {Aller}, {Geballe}, \&
  {Baas}}]{1981ApJ...249..592B}
{Beck}, S.~C., {Lacy}, J.~H., {Townes}, C.~H., {et~al.} 1981, \apj, 249, 592

\bibitem[{{Bloecker}(1995)}]{1995A&A...299..755B}
{Bloecker}, T. 1995, \aap, 299, 755

\bibitem[{{Buerger}(1973)}]{1973ApJ...180..817B}
{Buerger}, E.~G. 1973, \apj, 180, 817

\bibitem[{{Cristallo} {et~al.}(2009){Cristallo}, {Straniero}, {Gallino},
  {Piersanti}, {Dom{\'{\i}}nguez}, \& {Lederer}}]{2009ApJ...696..797C}
{Cristallo}, S., {Straniero}, O., {Gallino}, R., {et~al.} 2009, \apj, 696, 797

\bibitem[{{Ercolano} {et~al.}(2004){Ercolano}, {Wesson}, {Zhang}, {Barlow}, {De
  Marco}, {Rauch}, \& {Liu}}]{2004MNRAS.354..558E}
{Ercolano}, B., {Wesson}, R., {Zhang}, Y., {et~al.} 2004, \mnras, 354, 558

\bibitem[{{Escalante} \& {Morisset}(2005)}]{2005MNRAS.361..813E}
{Escalante}, V. \& {Morisset}, C. 2005, \mnras, 361, 813

\bibitem[{{Ferland} {et~al.}(1998){Ferland}, {Korista}, {Verner}, {Ferguson},
  {Kingdon}, \& {Verner}}]{1998PASP..110..761F}
{Ferland}, G.~J., {Korista}, K.~T., {Verner}, D.~A., {et~al.} 1998, \pasp, 110,
  761

\bibitem[{{Fitzpatrick}(1999)}]{1999PASP..111...63F}
{Fitzpatrick}, E.~L. 1999, \pasp, 111, 63

\bibitem[{{Garc{\'{\i}}a-Segura} {et~al.}(2006){Garc{\'{\i}}a-Segura},
  {L{\'o}pez}, {Steffen}, {Meaburn}, \& {Manchado}}]{2006ApJ...646L..61G}
{Garc{\'{\i}}a-Segura}, G., {L{\'o}pez}, J.~A., {Steffen}, W., {Meaburn}, J.,
  \& {Manchado}, A. 2006, \apjl, 646, L61

\bibitem[{{Gesicki} {et~al.}(1996){Gesicki}, {Acker}, \&
  {Szczerba}}]{1996A&A...309..907G}
{Gesicki}, K., {Acker}, A., \& {Szczerba}, R. 1996, \aap, 309, 907

\bibitem[{{Gesicki} {et~al.}(2006){Gesicki}, {Zijlstra}, {Acker}, {G{\'o}rny},
  {Gozdziewski}, \& {Walsh}}]{2006A&A...451..925G}
{Gesicki}, K., {Zijlstra}, A.~A., {Acker}, A., {et~al.} 2006, \aap, 451, 925

\bibitem[{{Gut\'{\i}errez-Moreno} {et~al.}(1985){Gut\'{\i}errez-Moreno},
  {Cortes}, \& {Moreno}}]{1985PASP...97..397G}
{Gut\'{\i}errez-Moreno}, A., {Cortes}, G., \& {Moreno}, H. 1985, \pasp, 97, 397

\bibitem[{{Guzm{\'a}n} {et~al.}(2009){Guzm{\'a}n}, {Loinard}, {G{\'o}mez}, \&
  {Morisset}}]{2009AJ....138...46G}
{Guzm{\'a}n}, L., {Loinard}, L., {G{\'o}mez}, Y., \& {Morisset}, C. 2009, \aj,
  138, 46

\bibitem[{{Handler} {et~al.}(1997){Handler}, {Mendez}, {Medupe}, {Costero},
  {Birch}, {Alvarez}, {Sullivan}, {Kurtz}, {Herrero}, {Guerrero}, {Ciardullo},
  \& {Breger}}]{1997A&A...320..125H}
{Handler}, G., {Mendez}, R.~H., {Medupe}, R., {et~al.} 1997, \aap, 320, 125

\bibitem[{{Henry} {et~al.}(2000){Henry}, {Kwitter}, \&
  {Bates}}]{2000ApJ...531..928H}
{Henry}, R.~B.~C., {Kwitter}, K.~B., \& {Bates}, J.~A. 2000, \apj, 531, 928

\bibitem[{{Hillier} {et~al.}(2003){Hillier}, {Lanz}, {Heap}, {Hubeny}, {Smith},
  {Evans}, {Lennon}, \& {Bouret}}]{2003ApJ...588.1039H}
{Hillier}, D.~J., {Lanz}, T., {Heap}, S.~R., {et~al.} 2003, \apj, 588, 1039

\bibitem[{{Hillier} \& {Miller}(1998)}]{HM98}
{Hillier}, D.~J. \& {Miller}, D.~L. 1998, \apj, 496, 407

\bibitem[{{Hyung} {et~al.}(1994){Hyung}, {Aller}, \&
  {Feibelman}}]{1994PASP..106..745H}
{Hyung}, S., {Aller}, L.~H., \& {Feibelman}, W.~A. 1994, \pasp, 106, 745

\bibitem[{{Karakas} {et~al.}(2009){Karakas}, {van Raai}, {Lugaro}, {Sterling},
  \& {Dinerstein}}]{2009ApJ...690.1130K}
{Karakas}, A.~I., {van Raai}, M.~A., {Lugaro}, M., {Sterling}, N.~C., \&
  {Dinerstein}, H.~L. 2009, \apj, 690, 1130

\bibitem[{{Keenan} {et~al.}(2003){Keenan}, {Aller}, {Exter}, {Hyung}, \&
  {Pollacco}}]{2003ApJ...584..385K}
{Keenan}, F.~P., {Aller}, L.~H., {Exter}, K.~M., {Hyung}, S., \& {Pollacco},
  D.~L. 2003, \apj, 584, 385

\bibitem[{{Krueger} \& {Czyzak}(1970)}]{1970RSPSA.318..531K}
{Krueger}, T.~K. \& {Czyzak}, S.~J. 1970, Royal Society of London Proceedings
  Series A, 318, 531

\bibitem[{{Kudritzki} {et~al.}(1997){Kudritzki}, {Mendez}, {Puls}, \&
  {McCarthy}}]{1997IAUS..180...64K}
{Kudritzki}, R.~P., {Mendez}, R.~H., {Puls}, J., \& {McCarthy}, J.~K. 1997, in
  IAU Symposium, Vol. 180, Planetary Nebulae, ed. H.~J. {Habing} \&
  H.~J.~G.~L.~M. {Lamers}, 64--+

\bibitem[{{Macfarlane} {et~al.}(1994){Macfarlane}, {Cohen}, \&
  {Wang}}]{1994ApJ...437..351M}
{Macfarlane}, J.~J., {Cohen}, D.~H., \& {Wang}, P. 1994, \apj, 437, 351

\bibitem[{{Marigo} {et~al.}(2003){Marigo}, {Bernard-Salas}, {Pottasch},
  {Tielens}, \& {Wesselius}}]{2003A&A...409..619M}
{Marigo}, P., {Bernard-Salas}, J., {Pottasch}, S.~R., {Tielens}, A.~G.~G.~M.,
  \& {Wesselius}, P.~R. 2003, \aap, 409, 619

\bibitem[{{Mendez} {et~al.}(1986){Mendez}, {Forte}, \&
  {Lopez}}]{1986RMxAA..13..119M}
{Mendez}, R.~H., {Forte}, J.~C., \& {Lopez}, R.~H. 1986, Revista Mexicana de
  Astronomia y Astrofisica, 13, 119

\bibitem[{{Mendez} {et~al.}(1988){Mendez}, {Kudritzki}, {Herrero}, {Husfeld},
  \& {Groth}}]{1988A&A...190..113M}
{Mendez}, R.~H., {Kudritzki}, R.~P., {Herrero}, A., {Husfeld}, D., \& {Groth},
  H.~G. 1988, \aap, 190, 113

\bibitem[{{Monteiro} {et~al.}(2005){Monteiro}, {Schwarz}, {Gruenwald},
  {Guenthner}, \& {Heathcote}}]{2005ApJ...620..321M}
{Monteiro}, H., {Schwarz}, H.~E., {Gruenwald}, R., {Guenthner}, K., \&
  {Heathcote}, S.~R. 2005, \apj, 620, 321

\bibitem[{{Morisset}(2006)}]{2006IAUS..234..467M}
{Morisset}, C. 2006, in IAU Symposium, Vol. 234, Planetary Nebulae in our
  Galaxy and Beyond, ed. M.~J. {Barlow} \& R.~H. {Mendez}, 467--468

\bibitem[{{Patriarchi} \& {Perinotto}(1995)}]{1995A&AS..110..353P}
{Patriarchi}, P. \& {Perinotto}, M. 1995, \aaps, 110, 353

\bibitem[{{Pauldrach} {et~al.}(2004){Pauldrach}, {Hoffmann}, \&
  {M{\'e}ndez}}]{2004A&A...419.1111P}
{Pauldrach}, A.~W.~A., {Hoffmann}, T.~L., \& {M{\'e}ndez}, R.~H. 2004, \aap,
  419, 1111

\bibitem[{{P{\'e}quignot} {et~al.}(2001){P{\'e}quignot}, {Ferland}, {Netzer},
  {Kallman}, {Ballantyne}, {Dumont}, {Ercolano}, {Harrington}, {Kraemer},
  {Morisset}, {Nayakshin}, {Rubin}, \& {Sutherland}}]{2001ASPC..247..533P}
{P{\'e}quignot}, D., {Ferland}, G., {Netzer}, H., {et~al.} 2001, in
  Astronomical Society of the Pacific Conference Series, Vol. 247,
  Spectroscopic Challenges of Photoionized Plasmas, ed. G.~{Ferland} \& D.~W.
  {Savin}, 533--+

\bibitem[{{Pottasch} {et~al.}(2004){Pottasch}, {Bernard-Salas}, {Beintema}, \&
  {Feibelman}}]{2004A&A...423..593P}
{Pottasch}, S.~R., {Bernard-Salas}, J., {Beintema}, D.~A., \& {Feibelman},
  W.~A. 2004, \aap, 423, 593

\bibitem[{{Prinja} {et~al.}(2007){Prinja}, {Hodges}, {Massa}, {Fullerton}, \&
  {Burnley}}]{2007MNRAS.382..299P}
{Prinja}, R.~K., {Hodges}, S.~E., {Massa}, D.~L., {Fullerton}, A.~W., \&
  {Burnley}, A.~W. 2007, \mnras, 382, 299

\bibitem[{{Richer} {et~al.}(2008){Richer}, {L{\'o}pez}, {Pereyra}, {Riesgo},
  {Garc{\'{\i}}a-D{\'{\i}}az}, \& {B{\'a}ez}}]{2008ApJ...689..203R}
{Richer}, M.~G., {L{\'o}pez}, J.~A., {Pereyra}, M., {et~al.} 2008, \apj, 689,
  203

\bibitem[{{Sch{\" o}nberner} {et~al.}(2005){Sch{\" o}nberner}, {Jacob},
  {Steffen}, {Perinotto}, {Corradi}, \& {Acker}}]{2005AA...431..963S}
{Sch{\" o}nberner}, D., {Jacob}, R., {Steffen}, M., {et~al.} 2005, \aap, 431,
  963

\bibitem[{{Schmutz} {et~al.}(1989){Schmutz}, {Hamann}, \&
  {Wessolowski}}]{1989A&A...210..236S}
{Schmutz}, W., {Hamann}, W.-R., \& {Wessolowski}, U. 1989, \aap, 210, 236

\bibitem[{{Sharpee} {et~al.}(2004){Sharpee}, {Baldwin}, \&
  {Williams}}]{2004ApJ...615..323S}
{Sharpee}, B., {Baldwin}, J.~A., \& {Williams}, R. 2004, \apj, 615, 323

\bibitem[{{Sharpee} {et~al.}(2003){Sharpee}, {Williams}, {Baldwin}, \& {van
  Hoof}}]{2003ApJS..149..157S}
{Sharpee}, B., {Williams}, R., {Baldwin}, J.~A., \& {van Hoof}, P.~A.~M. 2003,
  \apjs, 149, 157

\bibitem[{{Sterling} {et~al.}(2007){Sterling}, {Dinerstein}, \&
  {Kallman}}]{2007ApJS..169...37S}
{Sterling}, N.~C., {Dinerstein}, H.~L., \& {Kallman}, T.~R. 2007, \apjs, 169,
  37

\bibitem[{{Tayal}(2000)}]{2000ADNDT..76..191T}
{Tayal}, S.~S. 2000, Atomic Data and Nuclear Data Tables, 76, 191

\bibitem[{{Tayal}(2006)}]{2006ApJS..163..207T}
{Tayal}, S.~S. 2006, \apjs, 163, 207

\bibitem[{{Vassiliadis} \& {Wood}(1994)}]{1994ApJS...92..125V}
{Vassiliadis}, E. \& {Wood}, P.~R. 1994, \apjs, 92, 125

\bibitem[{{Wilson} \& {Bell}(2002)}]{2002MNRAS.331..389W}
{Wilson}, N.~J. \& {Bell}, K.~L. 2002, \mnras, 331, 389

\bibitem[{{Wright} {et~al.}(2007){Wright}, {Barlow}, {Ercolano}, \&
  {Rauch}}]{2007arXiv0709.2122W}
{Wright}, N.~J., {Barlow}, M.~J., {Ercolano}, B., \& {Rauch}, T. 2007, ArXiv
  e-prints, 709

\end{thebibliography}

\Online
\appendix

             \section{Presence of an X-ray component}
\label{sec:xray}
\begin{figure*}
  \centering
    \includegraphics[width=16.cm]{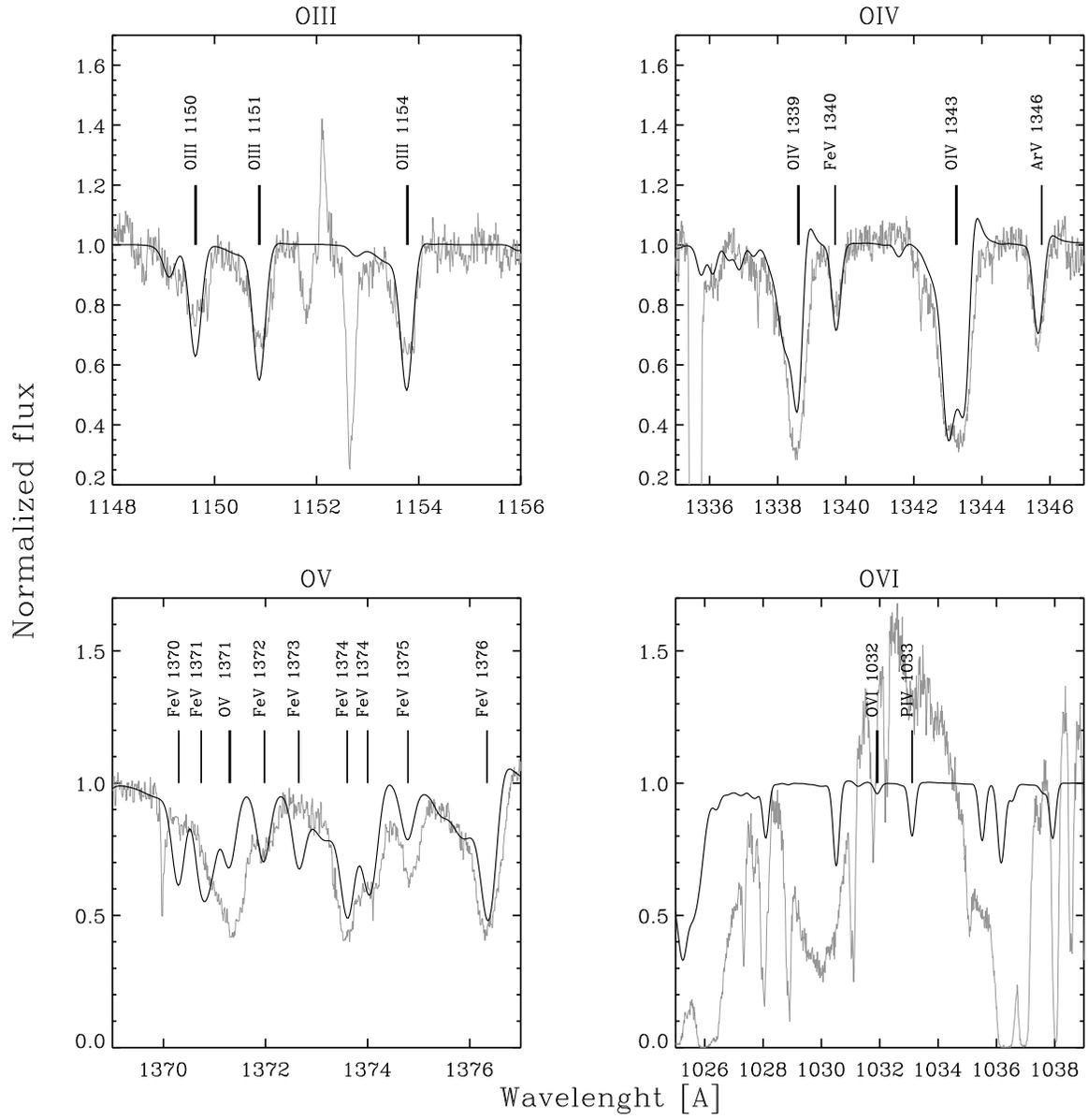}
   \caption{The fits to the oxygen lines. Argon and some FeV lines are also indicated.}
  \label{fig:oxy_fig}
\end{figure*}

The UV spectra exhibit a strong stellar \dalloa{O}{VI}{1031}{1037} line (Fig.~\ref{fig:oxy_fig}) that cannot be reproduced by a $\sim40$~kK stellar atmosphere model. This line is classically explained by the Auger ionization due to X-ray emission produced by wind instabilities. A weak X-ray emission is confirmed by Chandra observations\footnote{ID: 7440, exposure time of 50 ks, obtained the 12th of December 2006}. The Chandra data reduction has been performed using the Ciao V3.4 package CALDB (updated in May 2007) following standard procedures. Images in the 0.5-1.5~keV, 1.5-2.5~keV, and 2.5-8~keV bands were generated using ``dmcopy'' and ``csmooth'' Ciao procedures. Counts are only significantly detected in the softer energy band image (0.5-1.5~keV), indicating that the source is intrinsically soft. 

The limited count detection (less than 40 photons) does not allow a robust spectral analysis. The soft energy flux (i.e. 0.5-2~keV band) is estimated by assuming an emission model of a black-body with a temperature of $\sim$ 0.03~keV and taking into account a neutral absorption of N$_{\rm H}=2\times 10^{21} \rm{cm}^{-2}$. The unabsorbed measured flux of the source in the 0.5-2~keV band is $1 \times 10^{-14}\rm{erg/s/cm}^2$. 

To take this X-ray emission into account, we add a black-body emission at 300~kK to the stellar flux. We do not have any constraints on the shape of the X-ray SED emission, so we used the Planck function as the simplest SED. 

This soft X-ray component does not affect the determination of the nebular parameters, such as the abundances. It does have a small influence on the high ionization potential (IP) emission lines (\forb{Ne}{III} and \forb{Ar}{IV}), which are slightly underestimated by the model if the X-ray component is not taken into account. We adjust the luminosity of the soft X-ray component, so that these high IP emission lines are reproduced. The required bolometric luminosity is 3 L$_\odot$ (using 1.25~kpc for the distance), which corresponds to a flux in the 0.5-2~keV band of $4 \times 10^{-17}\rm{erg/s/cm}^2$, much lower than the observed X-ray flux. A hotter X-ray source would have a lower luminosity to conserve the soft X-ray flux and no influence on these lines. In contrast, a cooler X-ray source would require a higher luminosity, leading to intensities for these high IP lines that are higher than the observed values.

             \section{Line profiles}
\label{sec:line-profiles}
\citet{1996A&A...309..907G} published observations of high spectral resolution of IC418 for the \hbeta/, \forba{O}{iii}{5007}, and \forba{N}{ii}{6584} lines. \citet{2004ApJ...615..323S} also presented high resolution observations for various ions. In both cases, the \ion{H}{i} profiles are broad, with half width at half maximum (HWHM) close to 18km/s, the [\ion{O}{iii}] lines are narrower, with a HWHM close to 9 km/s, while the low density species, such as [\ion{N}{ii}], show a double-peaked profile, with an inter-peak spacing of 20 km/s and each peak having a HWHM comparable to the one of the single [\ion{O}{iii}] line. 

\begin{figure}
  \centering
    \includegraphics[width=9.cm]{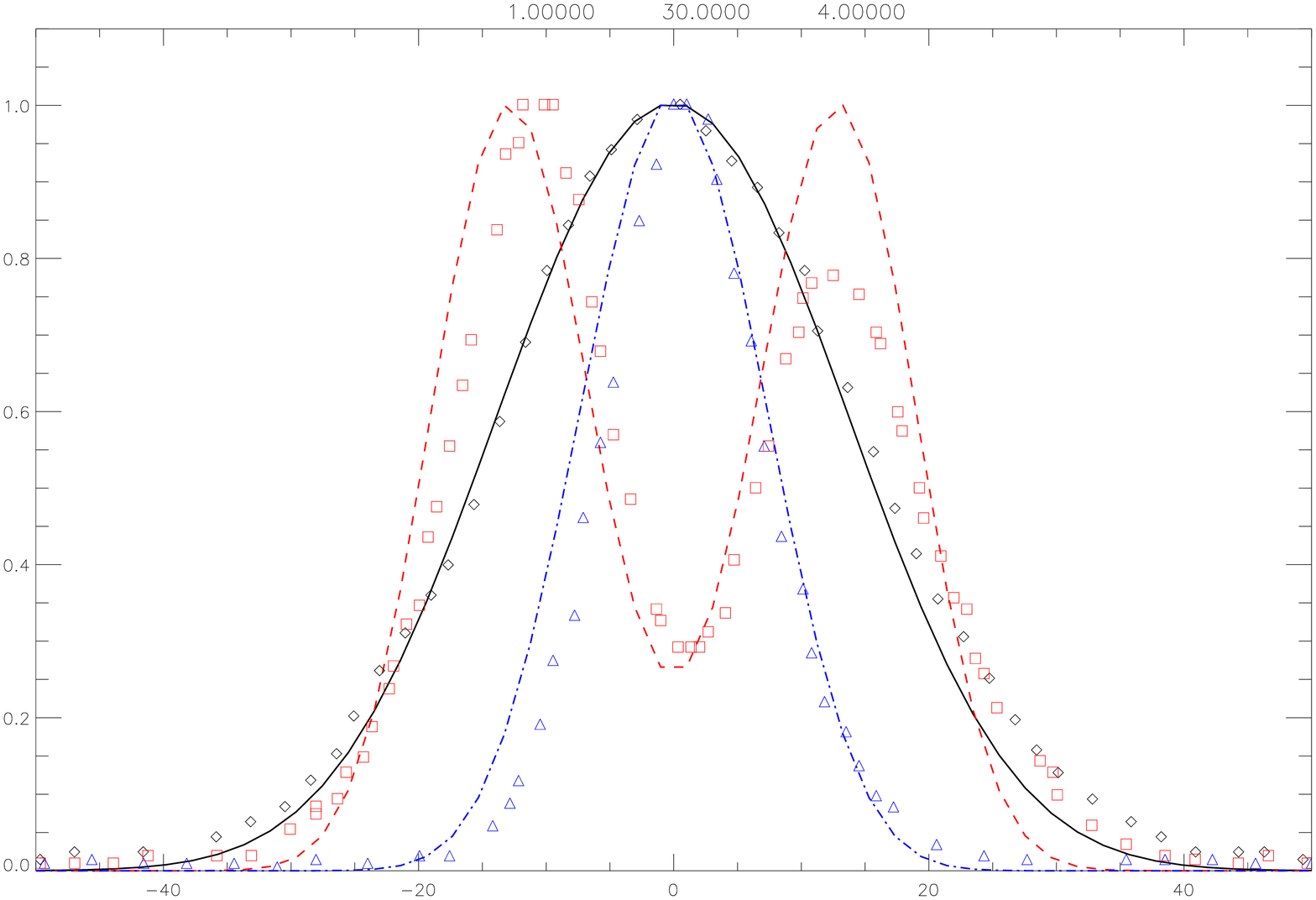}
  
  \caption{Emission-line profiles: models (lines) and observations \citep[][symbols]{1996A&A...309..907G}, for \hbeta/ (black solid line and diamonds), \forba{O}{iii}{5007} (blue dot-dashed line and triangles), and \forba{N}{ii}{6584} (red dashed line and squares). Intensities are scaled so that the maximum of each profile reaches 1.0.
  }
  \label{fig:line.profiles}
\end{figure}

We compute the line profiles and compare them to the observations of \citet{1996A&A...309..907G} using the same aperture size and position they used, namely a centered square aperture of 3 arcsec size. 
Despite the density law being determined by these authors in no way reproduces the observed surface brightness, we can fit an expansion velocity law similar to the one that they adopt. The resulting profiles reproduce the main properties of the observed profiles (see Fig.~\ref{fig:line.profiles}). The velocity law we adopt here is $V \propto R^4$ with a maximum of 30 km/s at the outer edge of the nebula.
It is beyond the scope of this paper to make a complete dynamical model of IC418. We only check that the morphology we used is not trivially ruled out by the line profile observations. The values for the velocities of O$^{++}$ and N$^+$ that we found are compatible with the radiation-hydrodynamic models presented by \citet{2005AA...431..963S}.

              \section{Chlorine collision strengths and electron temperature diagnostic}
\label{sec:chlor-coll-strengths}

The collision strengths for Cl$^{++}$ used in Cloudy are taken from \citet{2002MNRAS.331..389W}. Note that Cloudy takes into account the correction by a factor of 10 for the collision strength for $^1$D$_2$-$^1$S$_0$ as discussed in \citet{2003ApJ...584..385K}.
We test these values against the previous values from \citet{1970RSPSA.318..531K} and found that these latest are in closer agreement with the observations. On one side, the values for [\ion{Cl}{ii}] and [\ion{Cl}{iii}] are fitted when using values from \citeauthor{1970RSPSA.318..531K}, while it is impossible to fit both ions with the same Cl abundance when using collision strengths from  \citeauthor{2002MNRAS.331..389W}. In this case, the  [\ion{Cl}{ii}] lines are predicted to be too high, while the [\ion{Cl}{iii}] lines are fitted well.

On the other hand, the electron temperature diagnostic \rforba{Cl}{ii}{6162}{9124} is found to be predicted coherent with the [\ion{S}{ii}] and one of the two  [\ion{O}{ii}] diagnostic when using the data of \citeauthor{1970RSPSA.318..531K}. For these low ionization lines, the model overpredicts the ratios with  $\kappa([\ion{S}{ii}], [\ion{Cl}{ii}] ) \sim 1.5$. 
Using the values from \citeauthor{2002MNRAS.331..389W} leads to a prediction of the  \rforba{Cl}{ii}{6162}{9124} ratio that is lower than what is observed, with $\kappa([\ion{Cl}{ii}]) \sim -1.2$.

\section{Online figures and tables}

\begin{figure*}
  \centering
    \includegraphics[width=16.cm]{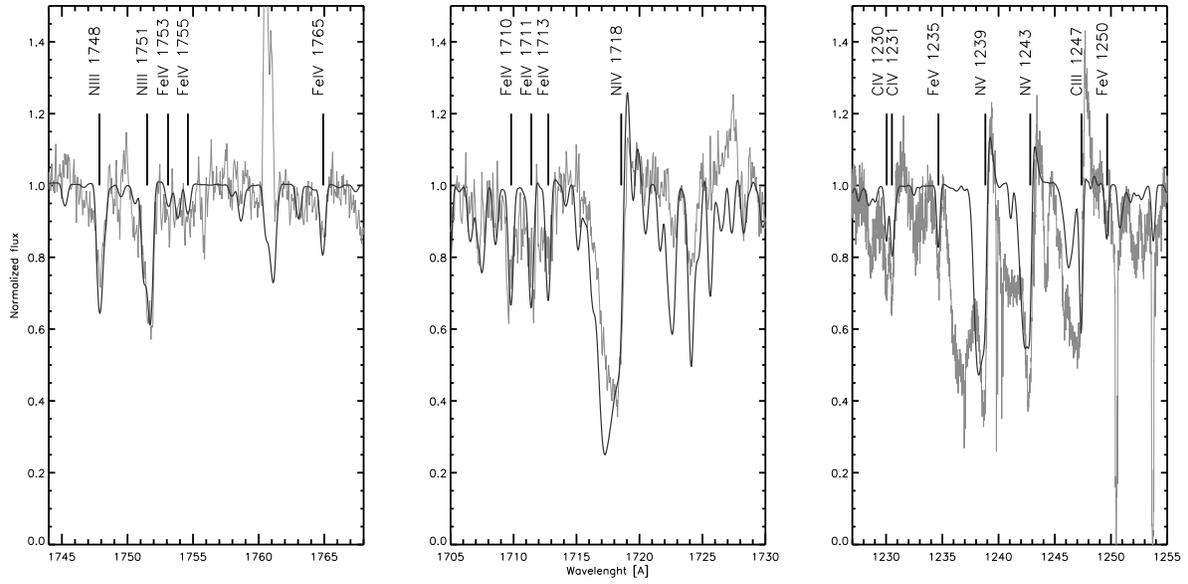}
   \caption{The fits to the nitrogen and FeIV lines. Note the nebular emission line \alloa{C}{II}{1761}.}
  \label{fig:nit_fig}
\end{figure*}

\begin{figure*}
  \centering
    \includegraphics[width=16.0cm]{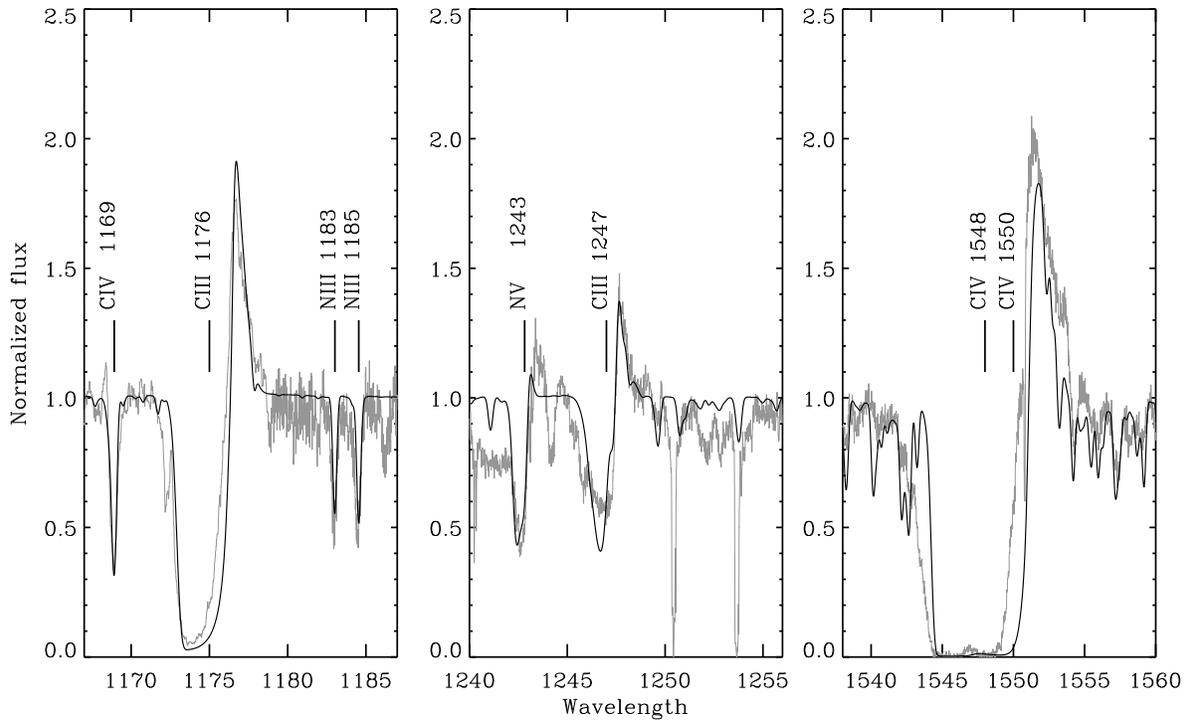}
   \caption{The fits to the carbon lines. \dalloa{N}{IV}{1183}{85} and \alloa{N}{V}{1188} lines are also marked.}
  \label{fig:carb_fig}
\end{figure*}

\begin{figure*}
  \centering
    \includegraphics[width=16.0cm]{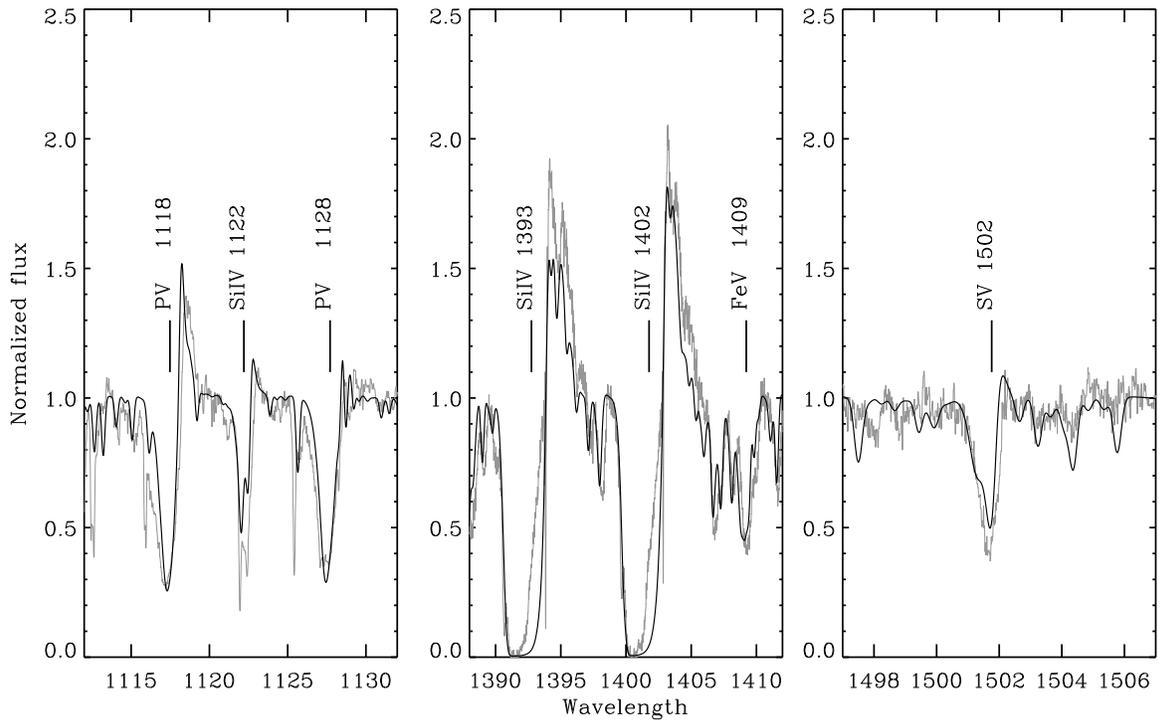}
   \caption{Phosphorus and silicon lines. Note the strong absorption lines.}
  \label{fig:phos_fig}
\end{figure*}


\begin{figure*}
  \centering
    \includegraphics[width=8.cm]{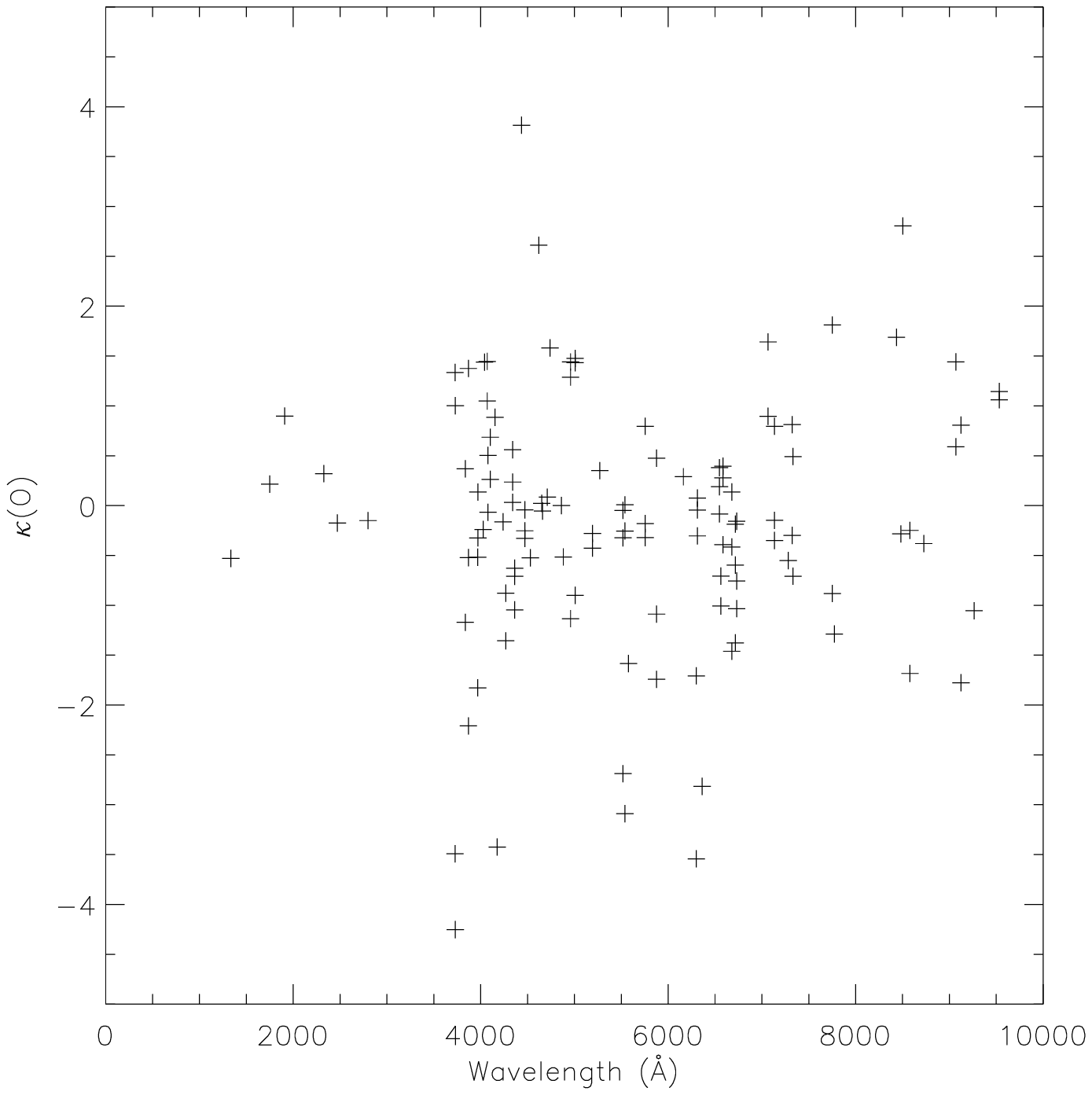}
    \includegraphics[width=8.cm]{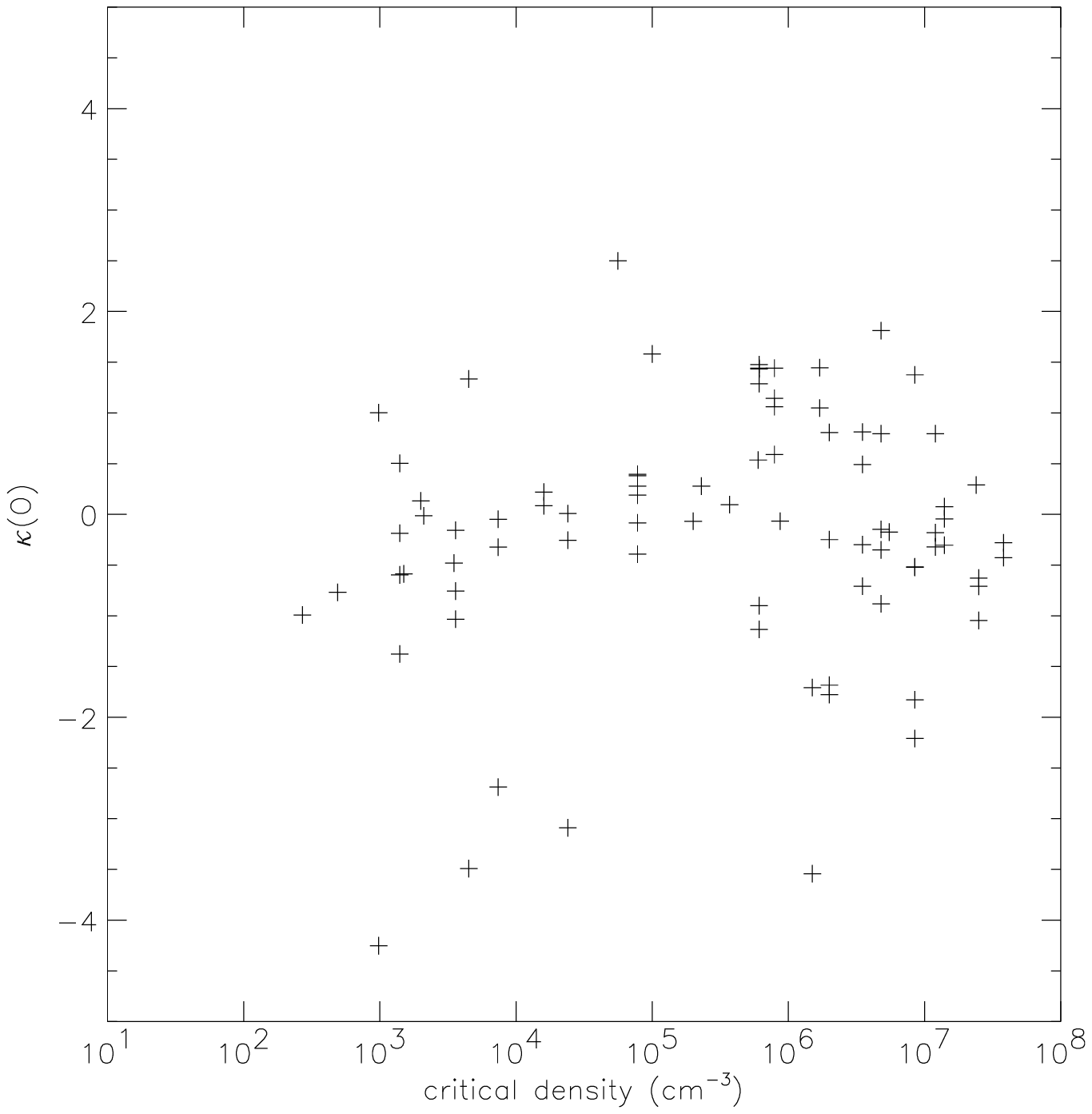}
  \caption{Quality factors $\kappa({\rm O})$  versus the wavelength (upper panel) and critical density (bottom panel).
  }
  \label{fig:reslambda}
\end{figure*}

\begin{figure*}
  \centering
    \includegraphics[width=8.cm]{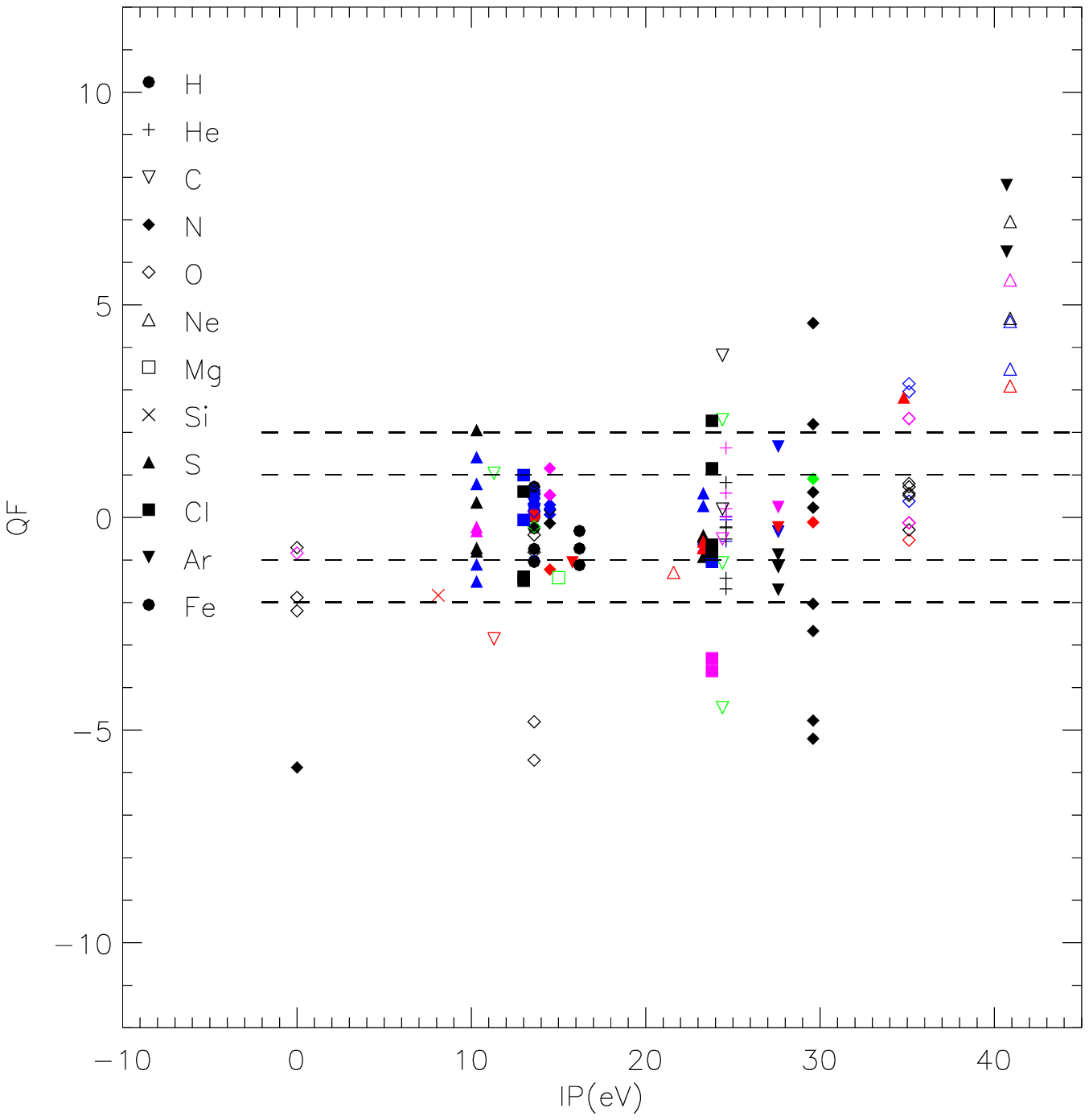}
  \caption{Same as Fig.~\ref{fig:QF.IP}, but for a black-body model. }
  \label{fig:QF.IP-BB}
\end{figure*}

\begin{figure*}
  \centering
    \includegraphics[width=8.cm]{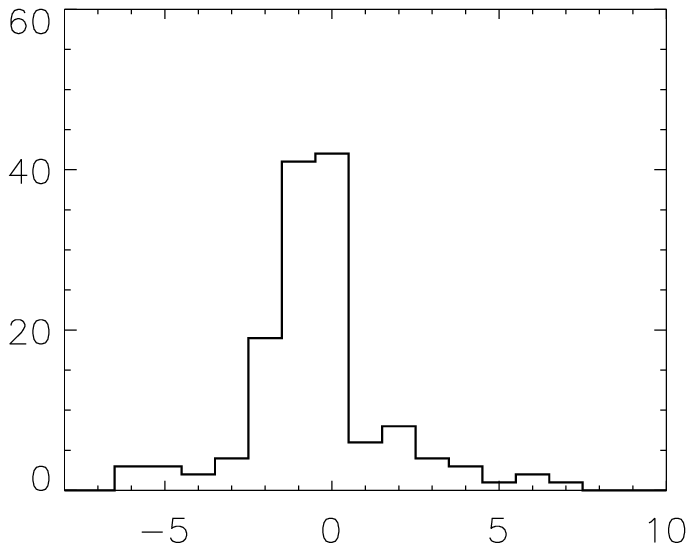}
  
  \caption{Same as Fig.~\ref{fig:histo}, but for black-body model. }
  \label{fig:histo-BB}
\end{figure*}

\begin{figure*}
  \centering
    \includegraphics[width=7.cm]{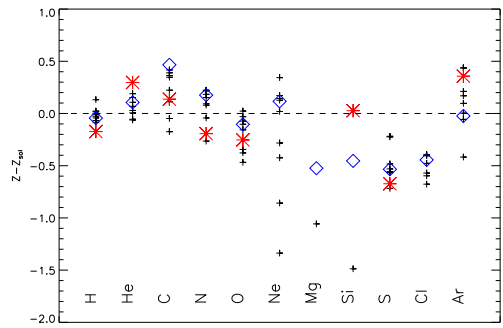} 
  \caption{Nebular abundances by mass from the literature (small crosses, same as in Table~\ref{tab:abund}) and from the adopted model (large stars represent the stellar abundance, and large diamond the nebular abundances). Units are in log(abundance by number) relative to the solar values from \citet{2005ASPC..336...25A}.}
  \label{fig:abund}
\end{figure*}

\begin{table}
  \caption{Diagnostic line ratios, see text for the columns definitions.} \label{tab:diag1} 
\centering
  \begin{tabular}{|l|r|r|r|}
    \hline
    \multicolumn{1}{|c|}{\textbf{I}} 
    & \multicolumn{1}{c|}{\textbf{II}}
    & \multicolumn{1}{c|}{\textbf{III}}
    & \multicolumn{1}{c|}{\textbf{IV}}\\
    \hline
R  O  2 2471/7325+     & {\bf     0.7635}&    0.7635&{\it      -0.00}\\
 Ne S  2 6731/6716      & {\bf     2.1250}&    2.0641&{\it      -0.12}\\
 Ne S  2 6731/6716 HAF  & {\bf     1.8914}&    2.0693&{\it       0.25}\\
 Ne S  2 6731/6716 GCM  & {\bf     2.0431}&    2.0588&{\it       0.02}\\
 Ne O  2 3726/3729      & {\bf     2.3653}&    2.5433&{\it       0.55}\\
 Ne O  2 3726/3729 HAF  & {\bf     2.4000}&    2.5500&{\it       0.24}\\
 Ne S  3 18.7/33.5      & {\bf     5.8102}&    6.1838&{\it       0.15}\\
 Ne Cl 3 5538/5518      & {\bf     1.9560}&    1.9852&{\it       0.04}\\
 Ne Cl 3 5538/5518 HAF  & {\bf     2.0000}&    2.0465&{\it       0.05}\\
 Ne Cl 3 5538/5518 GCM  & {\bf     2.2222}&    1.9420&{\it      -0.30}\\
 Ne O  3 51.9/88.4      & {\bf     6.3413}&    7.2426&{\it       0.33}\\
 Ne Ar 4 4740/4711      & {\bf     1.2000}&    1.7769&{\it       1.11}\\
 Te S  2 4068+/6731+    & {\bf     0.3915}&    0.5367&{\it       1.56}\\
 Te S  2 4068+/6731+HAF & {\bf     0.3198}&    0.5540&{\it       1.92}\\
 Te Cl 2 6162/9124+     & {\bf     0.0108}&    0.0182&{\it       1.60}\\
 Te O  2 7320+/3727+    & {\bf     0.1433}&    0.1949&{\it       3.10}\\
 Te O  2 7320+/3727+HAF & {\bf     0.2266}&    0.2042&{\it      -0.55}\\
 Te N  2 5755/6584+     & {\bf     0.0127}&    0.0124&{\it      -0.15}\\
 Te N  2 5755/6584+ HAF & {\bf     0.0141}&    0.0128&{\it      -0.32}\\
 Te S  3 6312/9531+     & {\bf     0.0143}&    0.0133&{\it      -0.25}\\
 Te S  3 6312/9531+ HAF & {\bf     0.0179}&    0.0141&{\it      -0.67}\\
 Te S  3 6312/18.7      & {\bf     0.0564}&    0.0543&{\it      -0.11}\\
 Te Ar 3 7136/8.99      & {\bf     1.1262}&    0.9620&{\it      -0.43}\\
 Te Ar 3 5192/7136+     & {\bf     0.0037}&    0.0036&{\it      -0.10}\\
 Te Ar 3 5192/7136+ HAF & {\bf     0.0043}&    0.0037&{\it      -0.39}\\
 Te N  3 1750/57.4      & {\bf     0.1757}&    0.2336&{\it       0.58}\\
 Te O  3 4363/5007+     & {\bf     0.0033}&    0.0027&{\it      -0.68}\\
 Te O  3 4363/5007+ HAF & {\bf     0.0045}&    0.0028&{\it      -1.32}\\
 Te O  3 4363/5007+ GCM & {\bf     0.0045}&    0.0027&{\it      -1.40}\\
 Te O  3 5007+/88.3     & {\bf   119.1307}&  140.8424&{\it       0.49}\\
 Te O  3 5007+/88.3 HAF & {\bf    47.9473}&   80.5872&{\it       1.45}\\
 Te Ne 3 3869+/15.5     & {\bf     0.4268}&    0.3934&{\it      -0.23}\\
 Te Ne 3 3869+/15.5 HAF & {\bf     0.2828}&    0.1607&{\it      -1.47}\\
    \hline
  \end{tabular}
\end{table}

\end{document}